\documentclass[pre,twocolumn,showpacs,superscriptaddress]{revtex4}
\usepackage{graphicx}
\usepackage{amsmath}
\usepackage{bm}
\usepackage{latexsym}
\begin{document}

\title{Metastable Intermediates in the Condensation of Semiflexible Polymers}

\author{B.~Schnurr}
\email{bernhard.schnurr@weizmann.ac.il} \affiliation{Department of
Physics \& Biophysics Research Division, University of Michigan,
Ann Arbor, MI 48109-1120} \affiliation{Department of Physics of
Complex Systems, Weizmann Institute of Science, Rehovot 76100,
Israel}
\author{F.~Gittes}
\email{gittes@wsu.edu} \affiliation{Department of Physics \&
Biophysics Research Division, University of Michigan, Ann Arbor,
MI 48109-1120} \affiliation{Department of Physics, Washington
State University, Pullman, WA 99164-2814}
\author{F.~C.~MacKintosh}
\email{fcm@nat.vu.nl} \affiliation{Department of Physics \&
Biophysics Research Division, University of Michigan, Ann Arbor,
MI 48109-1120} \affiliation{Division of Physics and Astronomy,
Vrije Universiteit, 1081 HV Amsterdam, The Netherlands}

\date{\today}

\begin{abstract}
Motivated by results from an earlier Brownian Dynamics (BD)
simulation for the collapse of a single, stiff polymer in a poor
solvent [B.~Schnurr, F.~C.~MacKintosh, and D.~R.~M.~Williams,
Europhys. Lett. \textbf{51}(3), 279 (2000)] we calculate the
conformational energies of the intermediate (racquet) states
suggested by the simulations. In the absence of thermal
fluctuations (at zero temperature) the annealed shapes of these
intermediates are well-defined in certain limits, with their major
structural elements given by a particular case of Euler's
elastica. In appropriate units, a diagram emerges which displays
the relative stability of all states, tori and racquets. We
conclude that, in marked contrast to the collapse of flexible
polymers, the condensation of semiflexible or stiff polymers
generically proceeds via a cascade through metastable
intermediates, the racquets, towards a ground state, the torus or
ring, as seen in the dynamical simulations.
\end{abstract}

% 87.15.He (Dynamics and conformational changes)
% 36.20.Ey (Conformation (statistics and dynamics))
% 87.15.-v (Biomolecules: structure and physical properties)

\maketitle
%\tableofcontents

\section{Introduction}

The conformation of individual polymer chains depends on the
properties of their environment, \textit{i.e.} the
solvent~\cite{degennes1979scaling, doiedwards1988,
grosberg1994statphys}. In the presence of a \textit{poor} solvent,
isolated polymer chains tend to collapse toward compact states, in
which polymer-solvent contacts are minimized. For flexible
polymers, the kinetics of this coil-globule transition have been
the subject of much research over the past few
decades~\cite{degennes1985kin, ostrovsky1994, buguin1996,
dawson1997, halperin2000, lee2001}. The kinetic pathway for
flexible polymer collapse has only recently been experimentally
confirmed to involve the formation of a pearl necklace and the
gradual diffusion of large pearls from the chain
ends~\cite{chu1995, abrams2001}.

In contrast to the flexible case, many polymers exhibit
substantial bending stiffness, thus adding the (opposing) tendency
to form extended structures. This makes a compact globule
energetically unfavorable for \textit{semiflexible} polymers
because compact globules involve large amounts of bending. Such
chains are described by the persistent or worm-like chain (WLC)
model~\cite{grosberg1994statphys}, examples of which include
predominantly biopolymers (\textit{e.g.} F-actin and DNA) but also
some synthetic polymers (\textit{e.g.} kevlar). The balance
between the tendencies to straighten the filament (due to a
bending energy) and to condense it (due to an effective
short-range attraction or poor solvent) is at the heart of the
condensation of semiflexible polymers.

The apparent equilibrium collapsed state for semiflexible polymers
is well-known: chains with significant bending stiffness can form
rings or toroids to avoid incurring the large bending penalty of a
spherical shell or a globule. This condensed state has been
suggested and studied theoretically~\cite{grosberg1979,
bloomfield1991, bloomfield1997, vasil1997}, demonstrated in a
variety of experimental systems~\cite{lifanding1992, fang1998surf,
shen2000, martin2000, liu2001interm, hansma2001surf}, and
confirmed by computer simulation~\cite{noguchi1996, byrne1998kin,
noguchi2000bd, stevens2001}. Theoretical work has predominantly
addressed structural features such as the detailed packing of
filaments~\cite{hud1995pha, hud1995rad, ubbink1995, ubbink1996def,
park1998, golo1998}, while dynamical simulations and Atomic Force
Microscopy (AFM) among others have increasingly focused on kinetic
aspects and condensation intermediates~\cite{fang1998mica, he2000,
noguchi2000bd, martin2000, liu2001interm}.

A particular set of recent dynamical simulations of isolated
chains~\cite{schnurr2000epl} has strongly suggested a possible
(and in fact generic) pathway for the collapse of semiflexible
polymers. These simulations showed not only the eventual formation
of tori from extended chains quenched in poor solvent but
demonstrated a series of long-lived, partially collapsed
intermediate states. Very similar chain morphologies (our racquet
states) also appear in other simulation work~\cite{noguchi2000bd}
and AFM studies of DNA condensation~\cite{martin2000}. Motivated
by these results, we develop and analyze a hierarchical family of
metastable racquet states. In particular, we demonstrate that
their relative conformational energies are consistent with the
role they play in the simulations: they form an energetically
driven cascade of increasingly compact conformations with sharp
transitions between them.

We begin in Section~\ref{sec:review} with a brief summary of the
dynamical simulation results~\cite{schnurr2000epl} which motivated
this analysis. Section~\ref{sec:intermediates} addresses the
morphology and evolution of the shapes to be analyzed in the
remainder of the paper. Our approach to calculating the surface
contributions to the conformational energies is developed in
Section~\ref{sec:tools}, followed by the two main sections
containing a detailed analysis of torus and racquet states
(Sections~\ref{sec:torus_states} and~\ref{sec:racquet_states}
respectively). Section~\ref{sec:racqvstori} finally compares their
relative stability and discusses the qualitative agreement with
the dynamical simulation results we set out to understand.

\section{\label{sec:review}Brief Review of the Dynamical Simulation Results}

The work described in Ref.~\cite{schnurr2000epl} applied a
standard Brownian Dynamics (BD) algorithm~\cite{doiedwards1988} to
a bead-and-spring model of a single polymer chain in the plane to
capture the most general features of a rather complex and
biologically important process, the condensation of DNA. The
technical details of that study are discussed
elsewhere~\cite{schnurr2000epl, schnurrthesis}. Here, we merely
sketch the gross features and the generic results that motivated
our work in this paper.

The dynamical evolution of a simulated chain followed a Langevin
equation of the form
%%%
\begin{equation}
  \xi\frac{dx_i}{dt} = -\frac{\partial U}{\partial x_i} + \eta_i(t) = F_{x_i}
\end{equation}
%%%
for each bead $i$, where $\xi$ is the coefficient of viscous drag
($\bm{F}_{\text{visc}}=\xi\bm{v}$) and $\eta$ the random noise.
Each bead is displaced by $\Delta x_i=(F_{x_i}/\xi)\Delta t$
during a time step $\Delta t$. The potential $U$ contains all
interactions internal to the chain, including the bending energy,
a short-range attractive interaction between beads (mimicking poor
solvent conditions) and a very stiff longitudinal compliance.
After thermalization of each chain, the solvent quality was
quenched at $t=0$.

Previous work~\cite{schnurr2000epl, schnurrthesis} showed the
typical dynamical evolution of a relatively short chain (a few
persistence lengths) as a progression through well-defined stages
identified by three types of conformations: extended chain with
thermal undulations, various racquet states (see
Figs.~\ref{fig:annealed} and~\ref{fig:schematic}), and the torus
or ring. We also pointed out that the end-to-end distance of the
filament as a function of time changes sharply with the
conformational transitions between states. It is important to note
that the described conformations persist in time, as seen by
quasi-plateaus in the end-to-end distance evolution, each lasting
for the considerable time of about $10^6$~BD steps, about one
tenth of the entire condensation event. We can roughly estimate
the correspondence between simulation steps and physical time for
a particular system. To do this, we express the link length as a
fraction of the persistence length and substitute for the local
drag coefficient, assuming the viscosity of water. For F-actin,
such an estimate suggests that an entire simulation of $10^7$~BD
steps models a filament for about 0.1~seconds. For DNA, this
interval corresponds to a fraction of a millisecond.

Temporal persistence of racquet structures was seen throughout the
simulations, suggesting that metastable intermediates are a
general feature in this collapse. Presumably, energy barriers
between intermediates are responsible for their local
(meta-)stability but we have not attempted to estimate their size.

\section{\label{sec:intermediates}Identification of Intermediate States}

\begin{figure}[hbtp]
  \centering
  \includegraphics[width=0.8\columnwidth]{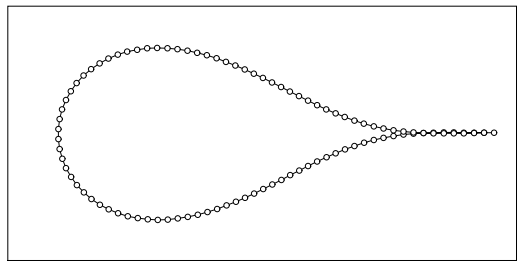}
  \caption{\label{fig:annealed}Annealed shape of a racquet head
  from the BD simulation, achieved by slowly lowering the
  effective temperature once the structure has formed. The shape
  coincides with the analytical curve to within a line width.}
\end{figure}

The dynamical simulations~\cite{schnurr2000epl} suggest that the
mechanism of collapse of semiflexible chains generically involves
transitions through a series of long-lived intermediate states. In
the absence of thermal fluctuations these intermediates anneal to
certain underlying shapes which are well-defined and allow a
straight-forward calculation of their conformational energies. The
crucial element in the underlying shapes is a characteristic
looped section which we call a racquet head. For the single
racquet, the shape of the head (see Fig.~\ref{fig:annealed}) was
produced in the simulation by annealing. While missing the effects
of thermal undulations, our calculations of the conformational
energies and detailed shapes of the annealed intermediates provide
an insightful framework for understanding the simulation results.

%%%%%%%%%%%%%%%%%%%%%%%%%%%%%%%%%%%%%%%%%%%%%%%%%%%%%%%%%%%%%%%%%%%%
%%%%%%%%%%%%%%%%%%%%%%%%%%%%%%%%%%%%%%%%%%%%%%%%%%%%%%%%%%%%%%%%%%%%

\begin{figure}[htbp]
  \centering
  \vspace{3mm}
  \includegraphics[width=0.7\columnwidth]{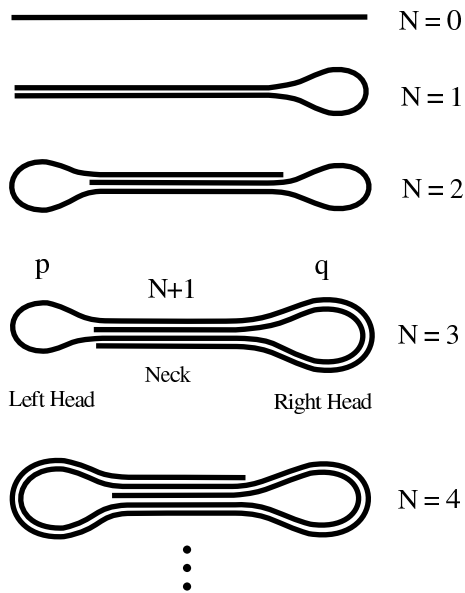}
  \caption{\label{fig:schematic}Schematic ``family'' of racquet
  states. The rod can be thought of as a trivial racquet without
  head ($N=0$). All subsequent states are labeled by their numbers
  of head sections.}
\end{figure}

In order to simplify the reference to specific states, we label
all racquet states by their number of looped sections. Thus the
rod is the $N=0$ state, the racquet with a single loop at one end
the $N=1$, and so forth, as indicated in Fig.~\ref{fig:schematic}.
We refer to the loop formed at the ends of the structure as the
``head'' and to the bundle of filaments connecting heads as the
``neck''. For the moment we neglect the more subtle question of
the exact location of filament ends. Naively one might assume that
filament ends coincide with the ends of the neck, since it is
straight; that is, the ends of the filament are expected to span
the entire neck as they incur no bending penalty, but generally
gain from increasing their overlap.

The picture as described provides an adequate starting point for
labeling the states we consider here. Among the shapes indicated
in Fig.~\ref{fig:schematic} we distinguish two basic racquet
symmetries: even and odd total numbers of heads $N$. For racquets
with even $N$, the number of overlaps in the head sections is
equal on the two sides, $p=q=N/2$. For racquets with odd $N$, one
side (we arbitrarily call it the left, following
Fig.~\ref{fig:schematic}) has one less $p=(N-1)/2$ than the other:
$q=(N+1)/2$. As a consequence, the filament ends of an even
racquet are on opposite sides of the neck.

Note that the dynamical simulations modeled the case of fixed
experimental conditions after the solvent quench. The polymer
chain is merely exploring a given conformational energy landscape
via thermal fluctuations. However, it can be instructive to
consider as a \textit{Gedankenexperiment} the case of variable
filament length, and we use this perspective in our discussion. A
(reduced) chain length is also a natural independent variable for
the presentation and comparison of states. In this alternative
perspective, the evolution of shapes starts with a short filament
that gradually lengthens. At first, only the neck grows until the
formation of a new head is favored. The incremental unit of growth
between conformations thus consists of one head plus one neck
segment. This procedure can be continued to arbitrary $N$ given
enough filament. In Section~\ref{sec:dimensionless} we will see
that the appropriate formulation of the problem accomplishes
changes in the (effective) filament length by adjusting the
solvent quality instead of the actual chain length.

%%%%%%%%%%%%%%%%%%%%%%%%%%%%%%%%%%%%%%%%%%%%%%%%%%%%%%%%%%%%%%%%%%%%
%%%%%%%%%%%%%%%%%%%%%%%%%%%%%%%%%%%%%%%%%%%%%%%%%%%%%%%%%%%%%%%%%%%%

\begin{figure}[htbp]
  \centering
  \includegraphics[width=\columnwidth]{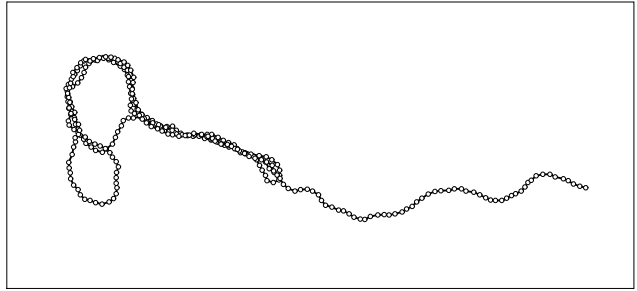}
  \caption{\label{fig:longchain}Early stages in the evolution of a
  long chain (300-mer, roughly $20\:\ell_{\text{p}}$) showing
  combinations of the conformational elements seen in shorter
  chains.}
\end{figure}

In the simulations, the actual transition into the torus state
could not be resolved in detail. It is clear however that there
are in principle at least two ways in which a loop can form: two
chain segments can meet with their tangent angles at an angle of
$\pi$ or $2\pi$. The former case leads to a racquet head, while
the latter makes a ring that allows the chain to wind up into a
torus directly. Since it is more likely for a stiff chain to bend
into the smaller angle, one would expect the transition to
racquets to be favored, at least for short chains. Statistically,
the simulations~\cite{schnurr2000epl} confirm this. Most of the
chains studied there were relatively short (less than 10
persistence lengths) though a few examples (see
Fig.~\ref{fig:longchain}) of longer chains showed qualitatively
similar behavior but with increased complexity, such as the
display of superstructures of racquets within racquets.

%%%%%%%%%%%%%%%%%%%%%%%%%%%%%%%%%%%%%%%%%%%%%%%%%%%%%%%%%%%%%%%%%%%%
%%%%%%%%%%%%%%%%%%%%%%%%%%%%%%%%%%%%%%%%%%%%%%%%%%%%%%%%%%%%%%%%%%%%

\section{\label{sec:tools}Conformational Energies}

Having discussed the basic morphology of the intermediate states,
we turn to the calculation of their conformational energies in the
absence of thermal undulations. Racquet and torus conformations at
zero temperature can be thought of as the underlying shapes, which
become modified by fluctuations at finite temperatures. Apart from
the bending contributions to the conformational energies, we need
to describe the nature of the surface energy by which we model the
poor solvent conditions that induce condensation. This surface
energy assumes the packing or bundling of filaments in a hexagonal
lattice (in cross section) and distinguishes between polymer and
solvent exposure. The local arrangement into a hexagonal columnar
phase has been confirmed for example by x-ray diffraction applied
to bundles of DNA and other charged polymer
chains~\cite{evdokimov1972, maniatis1974, evdokimov1976,
podgornik1989, podgornik1990} and the detailed structure within
bundles of semiflexible polymer chains has been studied
theoretically~\cite{odijk1993a, odijk1993b}.

Our calculations describe the simplified model with filaments of
vanishing thickness (compared to other length scales in the
problem), while we model their packing on a perfect hexagonal
lattice. Thus we do not take into account any winding defects due
to topological constraints or variations in curvature due to the
finite filament thickness. As implicit in the description of a
worm-like chain, we assume a uniform bending modulus. In the torus
state, such an ideal chain forms a circular ring with a single
radius of curvature. Furthermore, we assume for both racquet heads
and torus that partial filament overhang, an effective
non-uniformity in the bending modulus, does not change their shape
but only their size. Neglecting these higher order corrections is
certainly justified in the limit that the bundling number $N$ gets
large.

\subsection{\label{sec:hexagonal}Surface Energy of Hexagonal Bundles}

In a hexagonally close-packed bundle, each filament in cross
section can be thought of as having six sites occupied by either
solvent or polymer. The poor solvent lowers the energy for
polymer-polymer relative to polymer-solvent contacts. To express
the fact that there is a \textit{relative} energetic advantage for
filaments to bundle versus being exposed to solvent, we explicitly
evaluate the total number of solvent-exposed sites and express the
energy as a surface tension.

Particular surface energies are evaluated as follows. To find the
coordination number $\alpha_N$ for an $N$ bundle, consider the
total number of surfaces or binding sites in the bundle with
hexagonal order ($6N$). This number is proportional to the energy
of $N$ individual filaments completely exposed to solvent. To
account for the effect of bundling, we note that a bond
corresponds to the merging of two binding sites on neighboring
filaments. We thus subtract the number of bonds formed from only
half the number of sites ($3N$) to find the coordination number
$\alpha_N$. As an example of this numerology consider the cases
for $N=5$ and 10: for 5 filaments there are 7 bonds, resulting in
a coordination of 8, while 10 filaments make 19 bonds and thus
have a coordination of 11. Multiplying by the surface tension
parameter $\gamma$ finally yields the surface energy per unit
length for such a bundle.

\subsection{\label{subsec:magicnumbers}Filled Shells and ``Magic'' Numbers}

\begin{figure}[hbtp]
  \centering
  \includegraphics[width=\columnwidth]{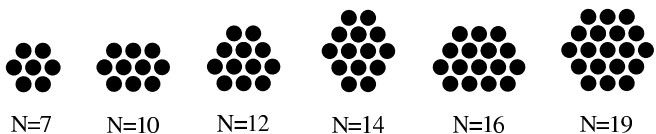}
  \caption{\label{fig:hexpack}Bundle cross sections of the lowest
  magic numbers on a hexagonal lattice. Note the arrangements of
  the``supermagic'' numbers 7 and 19 into perfect hexagons.}
\end{figure}

Differences between subsequent $\alpha_N$ are always either zero
or one (except between $N=1$ and 2). This creates non-uniformities
in the effective binding strengths per unit length, thus favoring
particular bundling numbers. We expect this effect for $N$ with
the same coordination as their predecessor ($\alpha_N =
\alpha_{N-1}$) but a coordination of one less than the $(N+1)$
bundle ($\alpha_N+1 = \alpha_{N+1}$). This is the case whenever an
added filament adds three instead of two bonds, thereby filling a
shell. Examples of this situation are found for
$N=7,10,12,14,16,19,21\ldots$ and we refer to them as ``magic''
numbers or filled shells. Cross sections of magic number bundles
correspond to arrangements with high degrees of symmetry, as shown
in Fig.~\ref{fig:hexpack}.

%%%%%%%%%%%%%%%%%%%%%%%%%%%%%%%%%%%%%%%%%%%%%%%%%%%%%%%%%%%%%%%%%%%%
%%%%%%%%%%%%%%%%%%%%%%%%%%%%%%%%%%%%%%%%%%%%%%%%%%%%%%%%%%%%%%%%%%%%

A few ``supermagic'' numbers ($N=7,19,37\ldots$) represent bundles
with the special symmetry of the equilateral hexagon; we will not
treat these cases separately. In agreement with Pereira and
Williams~\cite{pereira2000} we find that bundles of magic numbers
(and particularly the supermagic ones) play the role of preferred
states with increased stability.

\subsection{\label{sec:dimensionless}``Condensation'' Length and Energy}

The formulation of the problem as presented contains a
characteristic length scale which greatly simplifies the
discussion and presentation of our results. Balancing expressions
for typical bending and surface energies $(\kappa/L \sim \gamma
L)$ for a given filament length $L$ defines a measure we call the
\textit{condensation length} $L_c \equiv \sqrt{\kappa/\gamma}$.
Its role in the behavior of a chain under particular conditions is
the following: given the physical parameters $\kappa$ and $\gamma$
a filament much shorter than $L_c$ will rarely self-intersect and
therefore typically form an extended structure, while one much
longer than $L_c$ is likely dominated by overlaps and will form
collapsed or at least partially collapsed (intermediate)
structures.

Another combination of the two basic parameters $\kappa$ and
$\gamma$ sets an analogous energy scale, the \textit{condensation
energy} $U_c \equiv \sqrt{\kappa\gamma}$. With these measures, all
conformational energies $U_N$ can be presented in dimensionless
units, where physical energies and lengths are normalized by their
condensation values: $u_N \equiv U_N/U_c$ and $\lambda \equiv
L/L_c$. This formulation also provides a convenient (experimental)
realization of ``changing the filament length''. We can vary the
reduced length $\lambda$ by adjusting the values of $\kappa$ and
$\gamma$ independently.

\section{\label{sec:torus_states}Torus States}

\begin{figure}[hbtp]
  \centering
  \includegraphics[width=0.3\columnwidth]{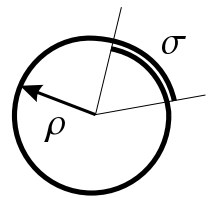}
  \caption{\label{fig:torussketch}Sketch of a generic torus (here
  a 1+ in our labeling scheme) of radius $\rho$ and overhang
  $\sigma$.}
\end{figure}

We describe a torus by the following two (dimensionless)
variables: a filament of length $\lambda$ is wound into a circle
of constant radius $\rho$, as shown in Fig.~\ref{fig:torussketch}.
In general, the torus can have any number $N$ of complete windings
(through an angle $2\pi$) and an amount of extra overhang $\sigma$
subject to the condition $\sigma < 2\pi\rho$. Since the entire
filament contour length $\lambda$ has a constant curvature, the
bending contribution to the conformational energy is always
$\lambda/2\rho^2$.

%%%%%%%%%%%%%%%%%%%%%%%%%%%%%%%%%%%%%%%%%%%%%%%%%%%%%%%%%%%%%%%%%%%%
%%%%%%%%%%%%%%%%%%%%%%%%%%%%%%%%%%%%%%%%%%%%%%%%%%%%%%%%%%%%%%%%%%%%

We find it convenient to distinguish forms of the torus with
different numbers of complete revolution $N$ defined as the
largest integer in $\lambda/2\pi\rho$. Any non-integer portion of
this ratio represents overhang of filament beyond complete
windings, defined as $\sigma \equiv \lambda-N(2\pi\rho)$. Our
distinction of different tori by $N$ naturally separates cases
with different coordination numbers and thus different surface
energies. In anticipation of an important distinction that
emerges, we call a torus without extra overhang ($\sigma=0$) an
``exact $N$'' while we refer to the generic torus with finite
overhang ($\sigma \neq 0$) as an ``$N+$''.

For the torus as described, we can then write down the following
expression for its total conformational energy
\begin{equation}
  u^{\text{torus}}_{N}(\lambda,\rho,\sigma) = \frac{\lambda}{2\rho^2}
  + 2\pi\rho\:[\alpha_N] + \sigma\:[\alpha_{N+1} - \alpha_N]
\end{equation}
with the common bending term followed by two surface terms
describing the contributions from the complete $N$-fold ring and the
extra piece of overhang $\sigma$, respectively.

Substituting for $\sigma$ leads to the torus energy in terms of
$\lambda$ and $\rho$ only, which allows us to find the equilibrium
size or radius $\rho_N(\lambda)$ for a particular state $N$ by
minimization with respect to $\rho$: note that $\partial^2
u^{\text{torus}}_{N}/\partial \rho^2 = 3\lambda/\rho^4$ is
positive everywhere. Resubstitution of $\rho_N(\lambda)$ yields
the conformational energy $u^{\text{torus}}_{N}(\lambda)$ in terms
of the single variable $\lambda$. The expressions found in this
way are valid in the ranges of $\lambda$ between $N$ and $N+1$
times the circumference $2\pi\rho_N$. However, a real solution for
this equilibrium size need not exist. This happens exactly for the
magic numbers with $N \geq 12$. When a real solution for the
equilibrium radius does exist, the resulting energy has two terms:
one proportional to $\lambda^{1/3}$, the other to $\lambda$. The
coefficient of the linear term is the combination of coordination
numbers ($\alpha_{N+1}-\alpha_N$) which can vanish for $N$ just
below the magic numbers (at $N=6,9,11,13,\ldots$) leaving these
cases with the functional dependence $\lambda^{1/3}$ only. Another
consequence of the numerology of hexagonal packing is that
different states $N$ can share the same energy expressions.
Examples are the series $N=(2,3,4,5)$ and the pairs
$N=(7,8),(17,18),(22,23),(25,26),(28,29)\ldots$\ These cases form
a particular class of transitions where $\sigma$ grows
continuously with $\lambda$.

\subsection{\label{subsec:stability}Stability}

\begin{figure}[hbtp]
  \centering
  \includegraphics[width=\columnwidth]{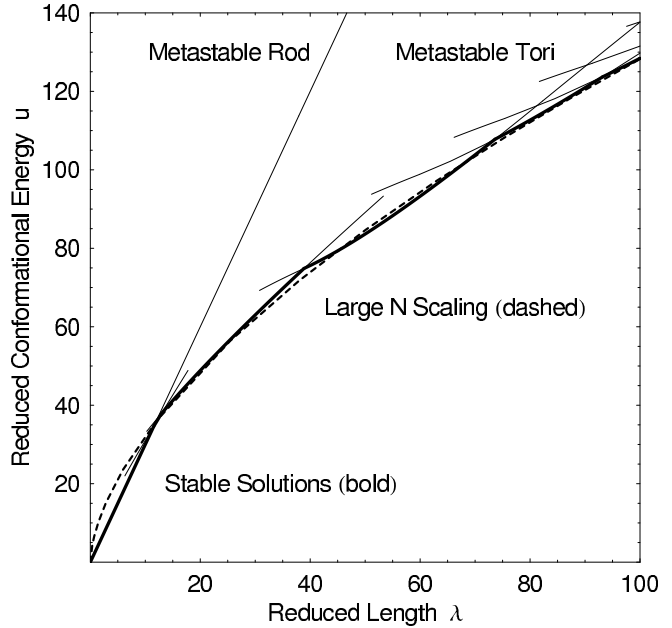}
  \caption{\label{fig:toroids}Conformational energies of the torus
  states as a function of filament length in reduced units. The
  thinner lines indicate (metastable) solutions in regions where
  they are not the ground state, which in turn is indicated by
  bold segments. The dashed line shows the large $N$ solution
  calculated in Section~\ref{subsec:torus_scaling}.}
\end{figure}

The above method for finding the optimal torus sizes by minimizing
the conformational energy represents the conventional approach to
determine metastability. However, there is a somewhat unusual
aspect to the problem at hand. The energy expressions for torus
states with different $N$ are in general not the same. This
introduces discontinuities in the form of the energy between
adjacent states. Consequently, there are not only the conventional
minima identified by their vanishing slopes but also another class
of solutions with discontinuities in slope at points where the
energy expressions to the left and right differ due to the
filament coordination. These are not minima in the usual sense
(for instance, they are not locally quadratic minima); they are
stabilized by finite slopes on both sides and do not have the
usual signature of a vanishing slope. Our results for the tori are
displayed in Fig.~\ref{fig:toroids}.

%%%%%%%%%%%%%%%%%%%%%%%%%%%%%%%%%%%%%%%%%%%%%%%%%%%%%%%%%%%%%%%%%%%%
%%%%%%%%%%%%%%%%%%%%%%%%%%%%%%%%%%%%%%%%%%%%%%%%%%%%%%%%%%%%%%%%%%%%

In order to establish the metastability of the tori of different
$N$ in more detail, we consider the behavior of the energy
derivatives with respect to the radius $\partial
u^{\text{torus}}_{N}/\partial \rho$ evaluated at the radii where
the exact $N$ and $N+1$ form. These derivatives are all monotonic
functions (functional dependence: $-\lambda^{-2}$) with at most a
single zero indicating a limit of metastability. Around these
zeroes, the derivative is generically negative to the left and
positive to the right. When negative, the energy is lowered by
increasing the radius $\rho$, thus driving any overhang $\sigma$
to vanish and making the torus an exact $N$. When positive, the
opposite is true, driving $\sigma$ to grow, making the torus an
$N+$. Note that there are cases (notably again for the magic
numbers with $N \geq 12$) where $\partial u_{N}/\partial \rho$ is
negative everywhere. These cases form an important class in which
tori never evolve (with increasing $\lambda$) towards states with
finite overhangs: they remain metastable (with complete or exact
overlap) for all lengths beyond some lower limit. The relative
positions of the zeroes in the energy derivatives combine in two
fundamental ways, resulting in exact and $N+$ tori for various
ranges of $\lambda$. A more detailed discussion of the various
cases can be found in~\cite{schnurrthesis}.

The rod ($N=0$) is of course a special (trivial) case without any
bending contribution. Due to the absence of any competition
between bending and self-affinity, the rod is (at least)
metastable for all lengths and follows a straight line. The lower
limit of the 1+ state is given by the circumference of a single
ring that just closes ($\lambda=2\pi$), a circle with the radius
of a condensation length. The subsequent small $N$ states show
variations depending on the numerology of the hexagonal
coordination numbers. For larger $N$ a perhaps generic type of
series emerges where $N+$ become exact $N+1$ which remain
metastable to infinity. Thus, in contrast to what we have
emphasized here by treating only cases with relatively small $N$
in detail, there appear to be only the few tori with $N$ below 12
that show variations to the generic pattern of exact (magic $N$)
tori without extra overhangs.

A direct comparison between relevant branches of the solutions
over regions of $\lambda$ provides the transition points between
stable (or ground) states, as summarized in
Table~\ref{tab:stable}.

\begingroup
\squeezetable
\begin{table}[htbp]
\caption{\label{tab:stable}Transition points for the lowest energy
states (ground states) up to $N=24$. Only for the ``shortest''
chains ($\lambda$ up to 11.543) is the rod stable to the torus.}
\begin{ruledtabular}
\begin{tabular}{cc}
State Labels & Transition Points \\
\colrule
0+ $\longrightarrow$ 1+ & 11.543 \\
1+ $\longrightarrow$ 2+ & 12.957 \\
2+ $\longrightarrow$ 3+ & 18.850 \\
3+ $\longrightarrow$ 4+ & 29.021 \\
4+ $\longrightarrow$ 7  & 38.871 \\
7  $\longrightarrow$ 10 & 73.625 \\
10 $\longrightarrow$ 12 & 93.195 \\
12 $\longrightarrow$ 14 & 119.876 \\
14 $\longrightarrow$ 16 & 148.687 \\
16 $\longrightarrow$ 19 & 155.672 \\
19 $\longrightarrow$ 24 & 228.700 \\
\vdots & \vdots \\
\end{tabular}
\end{ruledtabular}
\end{table}
\endgroup

It appears that the majority of stable torus states (perhaps all
$N\geq 12$) are exact states over their entire range. Their labels
$N$ are a subset of the magic numbers.

\subsection{Discreteness}

Another notable result are the discontinuities due to the
hexagonal packing and its discrete coordination numbers. We might
have expected the overhang $\sigma$ and the torus size $\rho$ to
be continuous with changes in $\lambda$. At least for small $N$
we find instead that small changes in $\lambda$ can cause discrete
jumps in the size of the ring $\rho$. This characteristic has
previously been described by Pereira and
Williams~\cite{pereira2000}. To what extent these effects are
experimentally observable is not known. As $N$ grows large, the
effect should weaken and ultimately disappear altogether.

%%%%%%%%%%%%%%%%%%%%%%%%%%%%%%%%%%%%%%%%%%%%%%%%%%%%%%%%%%%%%%%%%%%%
\begin{figure}[hbtp]
  \centering
  \includegraphics[width=\columnwidth]{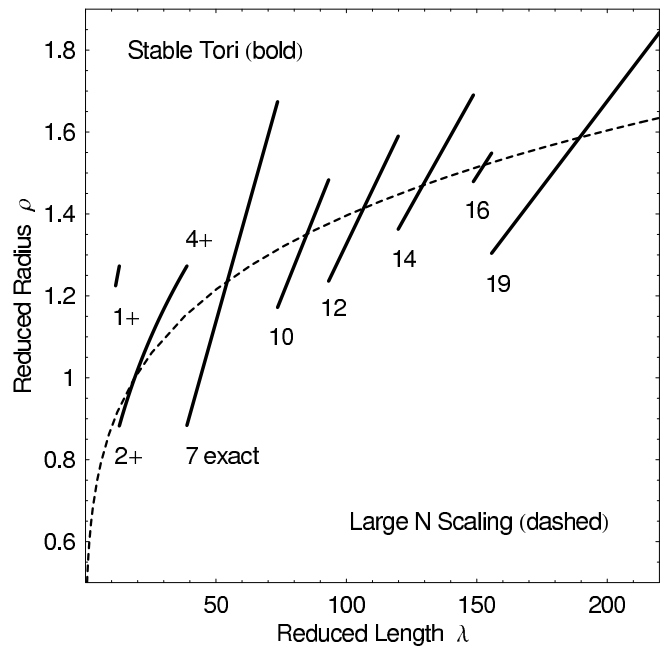}
  \caption{\label{fig:radii}Radii of the stable torus states
  (bold) as a function of filament length in reduced units,
  showing discrete transitions between the various series of
  states. The series shown are labeled by the states at their
  lower extremes. For comparison, the continuous large N solution
  is shown as a dashed line.}
\end{figure}
%%%%%%%%%%%%%%%%%%%%%%%%%%%%%%%%%%%%%%%%%%%%%%%%%%%%%%%%%%%%%%%%%%%%

Discrete jumps are perhaps most prominently displayed as
discontinuities in the torus size. Fig.~\ref{fig:radii} shows the
sizes or radii $\rho_N(\lambda)$ of the ground states as a
function of $\lambda$. The first bold curve segment starts where
the filament first makes a stable 1+. Note that the functional
dependence of the first two segments is different from the
subsequent (linear) ones. The first two series evolve continuously
according to their equilibrium solution for $\rho_N$ with a
functional dependence of $\lambda^{1/3}$. Their prefactors are
determined by some combination of the appropriate coordination
numbers. By contrast, all subsequent segments are due to solutions
which are constrained to be exact (by virtue of the magic numbers)
and thus have the linear dependence of $\lambda/2\pi N$. The
length of the various segments indicates the stability of the
states they represent. Clearly, states with supermagic bundling
numbers (the figure only shows $N=7$ and 19) are especially
stable. The dashed line indicates the (continuous) solution found
in the limit of large $N$, as discussed in the following section.

\subsection{\label{subsec:torus_scaling}Large $N$ Limit and Scaling}

The scaling argument of the torus size with filament length goes
back to Ubbink and Odijk~\cite{ubbink1995}. We sketch a similar
argument here in order to compare it in
Section~\ref{subsec:racquet_scaling} with the analogous argument
for the racquets. First we give the straightforward scaling
argument. In a second pass, we then determine the prefactors based
on the more accurate, hexagonally faceted cross section of the
torus.

In the limit as $N$ grows large, we can neglect such details as
partial overhangs (finite $\sigma$) since differences between $N$
and $N+1$ vanish as $1/N$. We assume first that the torus is a
perfect cylinder with circular cross section. It grows as $N$, the
number of filaments wound around its circumference. Thus, we
expect the total torus surface area to scale as $\rho\sqrt{N}$.
Substituting for the radius in a scaling sense
($\rho\sim\lambda/N$) we find that the conformational energy has
two terms: one proportional to $\lambda/N^{1/2}$, the other to
$N^2/\lambda$. Minimization with respect to $N$ (implicitly
letting $N$ be a continuous variable) yields the following set of
scaling relations.
\begin{subequations}
  \label{eq:scaling}
  \begin{eqnarray}
    N               &\sim& \lambda^{4/5} \\
    \rho_{\infty}   &\sim& \lambda^{1/5} \\
    u_{\infty}      &\sim& \lambda^{3/5}
  \end{eqnarray}
\end{subequations}
To find the prefactors, we need to consider a geometrically more
careful treatment. We assume that the torus formed has both a
perfect hexagonal cross section and an integer winding number (no
partial overhangs). These assumptions are reasonable: it can be
shown by direct calculation that the surface tension for a fixed
number of filaments on a triangular lattice is smaller for the
hexagonal than for the circular cross section. This is analogous
to a Wulff construction~\cite{wulff1901} which captures, for
instance, the faceting of crystals in solid state physics. The
integer winding number is justified since the difference in the
surface energies between $N$ and $N+1$ filaments vanishes as $N$
becomes very large.

For an (equilaterally) hexagonal cross section with its
symmetries, we can determine the following relationships
geometrically. As a characteristic for the size of the hexagon, we
label the integer number of lattice spacings on a side by $m$. The
counting of lattice sites (or filaments) in such a hexagonal
bundle is $N=3m^2$ to leading order in $m$. Proper counting adds a
linear and a constant term: $N=3m^2+3m+1$ but in the limit of
large $N$ we keep only the leading order in $m$.

We find the surface energy of such a bundle by counting
solvent-exposed filament sites. A filament at an edge (of which
there are $m-1$) exposes 2 sites, while one at a corner exposes 3.
Taking into account the 6-fold symmetry of the hexagon, this
results in $12m+6$ exposed sites on the surface. Substitution then
yields the limiting coordination number (a surface energy per unit
length) $\alpha_\infty = 2\sqrt{3\:N}$ for a hexagonal bundle of
$N$ filaments. This coordination number also provides the
prefactors for the conformational energy of the torus in the limit
of large $N$.
\begin{equation}
  u_\infty = \frac{2\sqrt{3}\:\lambda}{\sqrt{N}}
           + \frac{2\pi^2N^2}{\lambda}
\end{equation}
The expressions analogous to Eqs.~\ref{eq:scaling} with
geometrical prefactors are then
\begin{subequations}
  \label{eq:torus_prefactors}
  \begin{eqnarray}
    N              &=& \frac{3^{1/5}}{(2\pi)^{4/5}}\:\lambda^{4/5}
                       \approx 0.286\:\lambda^{4/5} \\
    \rho_{\infty}  &=& (6\pi)^{-1/5}\:\lambda^{1/5}
                       \approx 0.556\:\lambda^{1/5} \\
    u_{\infty}     &=& \frac{5(3\pi)^{2/5}}{2^{3/5}}\:\lambda^{3/5}
                       \approx 8.092\:\lambda^{3/5}.
  \end{eqnarray}
\end{subequations}
The last two expressions are shown in Figs.~\ref{fig:toroids} and
\ref{fig:radii} as dashed lines. We see outstanding agreement
between the large $N$ limit and the exact solutions down to the
lowest $N$ in Fig.~\ref{fig:toroids}.

\section{\label{sec:racquet_states}Racquet States}

\begin{figure}[htbp]
  \centering
  \includegraphics[width=0.8\columnwidth]{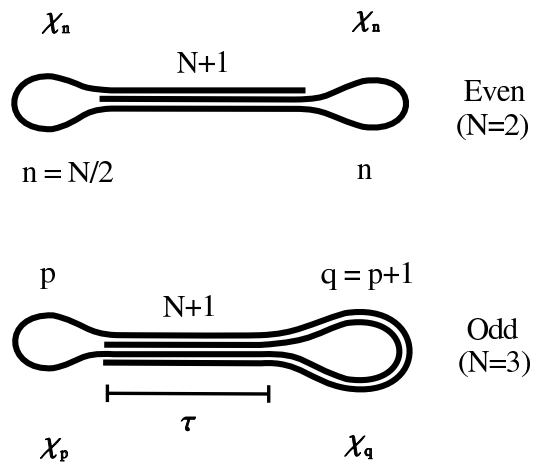}
  \caption{\label{fig:evenodd}Comparison of the structure and
  labels of generic even and odd racquets, represented here by the
  $N=2$ and 3. All the heads on one side are identical; the
  schematic separates head and neck filaments only to indicate
  their multiplicity.}
\end{figure}

The racquet conformational energies are made up of bending
contributions from each of the heads, and surface contributions
from the heads as well as the neck region in between. As shown in
Fig.~\ref{fig:evenodd}, the racquets divide naturally into two
groups: those with even and odd numbers of heads.

%%%%%%%%%%%%%%%%%%%%%%%%%%%%%%%%%%%%%%%%%%%%%%%%%%%%%%%%%%%%%%%%%%%%
%%%%%%%%%%%%%%%%%%%%%%%%%%%%%%%%%%%%%%%%%%%%%%%%%%%%%%%%%%%%%%%%%%%%

In the even case, the number of heads on each side equals $n
\equiv N/2$ by symmetry. In the odd case, we have $p \equiv
(N-1)/2$ heads on the ``left'' and $q \equiv (N+1)/2$ on the
``right''. The labels ``left'' and ``right'' are our arbitrary
naming convention (see Fig.~\ref{fig:evenodd}). The variables $p$
and $q$ for the bundling numbers of the heads always differ by 1
($q=p+1$) and sum to $N$. The bundling number of filaments forming
the neck is always $N+1$.

Given these bundling numbers, the remaining variables (in
dimensionless units) for the generic racquet are the overall
filament length $\lambda$, and the head sizes on the two sides
(namely the contour lengths of the heads, labeled $\chi_p$ and
$\chi_q$). So far we have described the racquets with their
filament ends coinciding with the ends of the neck. However, in
general (and in analogy with the overhang $\sigma$ in the torus
case) we need to allow for the extension of these ends into the
heads, or the retraction into the neck. Lengths of overhang are
labeled $\sigma_p$ and $\sigma_q$ and their sign indicates whether
they extend into or retract back from the heads. The length of the
neck $\tau$ is not an independent variable once all other
parameters are fixed, since the total filament length imposes a
constraint.

For the even racquet with a given $\lambda$, the number of
variables reduces to only two. Since the left and right heads for
even racquets are identical by symmetry, we collapse their labels
and are left with only one head size ($\chi_n \equiv \chi_p =
\chi_q$) and a single overhang variable ($\sigma_n \equiv \sigma_p
= \sigma_q$). The overall filament length for the even racquet is
distributed into $\lambda_N^{\text{even}} = N\chi_n + 2\sigma_n +
(N+1)\tau$ where the terms are ordered as heads, overhang, and
neck. For the odd racquet, we leave the left and right head sizes
separate, but require that any overhang be symmetrically
distributed on the left side. This is not the only possible
metastable solution, but the one we describe here generically as
the most symmetrical; we discuss the details of other possible
solutions further in Section~\ref{subsec:evenodd}. In the odd
case, the overall chain length divides itself into
$\lambda_N^{\text{odd}} = p\chi_p + q\chi_q + 2\sigma_p +
(N+1)\tau$ where we use the single overhang variable $\sigma_p$ to
indicate that the two possible pieces of overhang are always on
the left side (see Fig.~\ref{fig:evenodd}).

%%%%%%%%%%%%%%%%%%%%%%%%%%%%%%%%%%%%%%%%%%%%%%%%%%%%%%%%%%%%%%%%%%%%
\begin{figure}[hbtp]
  \centering
  \includegraphics[width=\columnwidth]{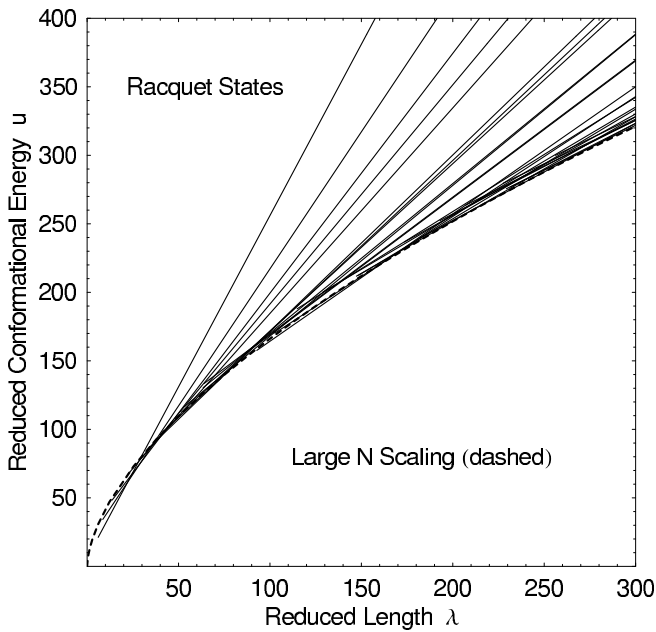}
  \caption{\label{fig:rac_scaling}Racquet state energies, shown as
  a function of filament length in dimensionless units, form a
  dense spectrum of solutions, each increasing linearly. For
  comparison, the scaling solution with proper prefactor in the
  limit of large $N$ is superimposed (dashed line). Note the
  nearly perfect agreement of the scaling solution with the lower
  envelope of the racquet states, down to the very lowest values
  of $N$.}
\end{figure}
%%%%%%%%%%%%%%%%%%%%%%%%%%%%%%%%%%%%%%%%%%%%%%%%%%%%%%%%%%%%%%%%%%%%

By way of a preview, we state here that the racquet solutions (see
Fig.~\ref{fig:rac_scaling}) differ fundamentally from those of the
tori. While the size of the torus was found to increase as a
function of $\lambda$ (up to discontinuities), the sizes of the
racquet heads (as well as any lengths of overlap) are fixed for
each state by the local force balance between the bundles of
filaments making up head and neck. Having determined the head
sizes and overhangs for a particular state, its lower limit of
validity $\lambda_{\text{low}}$ is found by adding up the head
sizes and overhangs in the absence of any neck at all ($\tau
\rightarrow 0$). This is the minimal filament length required to
form a particular racquet. For all lengths beyond, the racquets
remain metastable as their energies increase linearly with slope
$\alpha_{N+1}/(N+1)$. Adding extra filament to any racquet
configuration only lengthens its neck, while the head sizes and
any overhang remain fixed. As a consequence, all racquets are (at
least) metastable solutions for any $\lambda$ beyond their lower
cutoff $\lambda_{\text{low}}$. What remains to formulate the total
conformational energies of the racquets is the bending
contribution due to partial overhangs.

\subsection{\label{sec:elastica}Head Shape---an Elastica}

\begin{figure}[hbtp]
  \centering
  \includegraphics[width=0.8\columnwidth]{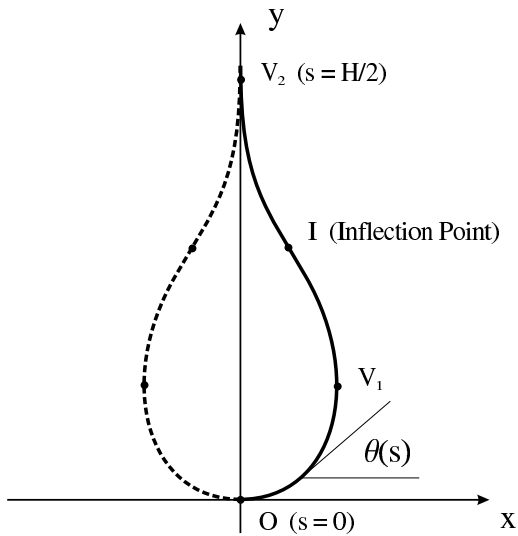}
  \caption{\label{fig:head}Schematic figure of a racquet head with
  axes appropriate for our calculation. Local tangent angles
  $\theta$ are measured from the $x$ axis. Symmetrical regions
  along the contour $s$ are delimited by solid circles.}
\end{figure}

Having identified the racquet head as the distinguishing common
element among the intermediate states, we calculate its
geometrical shape (see Fig.~\ref{fig:head}) in the absence of
thermal fluctuations from the bending of a slender, elastic rod.
The expression for this head shape is necessary for the
determination of bending energies for the racquets. The general
class of shapes resulting from the bending of a slender rod by
forces and couples applied at its ends only are known as
\textit{elastica}. Such solutions were first studied by Euler in
1744. The particular solution we seek is schematically drawn in
Fig.~51 of the treatise by Love~\cite{love1944treatise}.

%%%%%%%%%%%%%%%%%%%%%%%%%%%%%%%%%%%%%%%%%%%%%%%%%%%%%%%%%%%%%%%%%%%%
%%%%%%%%%%%%%%%%%%%%%%%%%%%%%%%%%%%%%%%%%%%%%%%%%%%%%%%%%%%%%%%%%%%%

To solve for the shape of a racquet head of total contour length
$H$ we consider the geometry as shown and labeled in
Fig.~\ref{fig:head}. In this section we use physical variables
instead of the dimensionless units introduced previously, as they
are more intuitive here and allow for dimensional analysis. Given
the obvious symmetry about the $y$ axis, it is sufficient to solve
for one half of the racquet head only. The tangent angle along the
curve increases from $\theta=0$ at the origin $O$ ($s=0$), via a
maximum at the inflection point $I$, to $\theta=\pi/2$ at $V_2$
($s=H/2$, where the head joins the neck). Note that there are two
points $V_1$ and $V_2$ at which the tangent is vertical
($\theta=\pi/2$), with an inflection point $I$ between them. These
three points define an additional symmetry (about the inflection
point $I$) for the contour between points $V_1$ and $V_2$.

Our particular elastica is solved~\cite{gittesthesis} by
minimizing the WLC (worm-like chain) Hamiltonian subject to the
boundary condition that the two halves of the head join in the
neck at $x=0$. We impose this constraint by means of a Lagrange
multiplier $\zeta$.
\begin{equation}
  U = \int_0^{\frac{H}{2}}ds
      \left\{ \frac{\kappa}{2} \left( \frac{\partial\theta}{\partial s} \right)^2
    + \zeta\cos\theta \right\}
\end{equation}
Applying Euler's equation to this expression leads to the
differential equation
\begin{equation}
  \label{eq:diffeq}
  \frac{d^2\theta}{ds^2} = -\beta^2 \sin\theta(s)
\end{equation}
where we made the substitution $\beta^2 \equiv \zeta/\kappa$ and
expressed the angle $\theta(s)$ explicitly as a function of the
contour length $s$. Note that the Lagrange multiplier $\zeta$ has
the dimensions of a force and expresses the force required to join
the two filament bundles in the neck. Eq.~\ref{eq:diffeq} can be
integrated to yield an expression for the curvature along the head
contour $s$ as a function of the tangent angle $\theta$.
\begin{equation}
  \label{eq:curvature}
  \frac{d\theta}{ds} = \frac{2\beta}{k} \sqrt{1-k^2\sin^2(\theta/2)}
\end{equation}
Equivalently, one can rearrange terms and express the contour
length $s$ scaled by $\beta$ in the form of an incomplete
elliptic integral of the first kind $F(\phi,k)$
\begin{equation}
  \label{eq:arclength}
  \beta s = k \int_0^{\frac{\theta}{2}}
            \frac{dt}{\sqrt{1-k^2\sin^2(t)}} = k\;F(\theta/2,k)
\end{equation}
where $k$ is the elliptic modulus (yet to be determined) and $t$
an integration variable. This is the parametric solution of an
elastica: it gives the contour length $s$ as a function of the
tangent angle $\theta$. The expression is multivalued over its
range, but invertible in certain regions. Four such regions are
defined by the axial symmetry of the head (through the origin and
the neck) and the pair of inflection points in between. We thus
cover the entire racquet head in a piecewise fashion, while only
two of these regions are essentially different: the piece from $O$
to $V_1$ and that from $V_1$ to $I$ (with its reflection from $I$
to $V_2$). The yet unknown elliptic modulus $k$ for our elastica
is found from a geometrical constraint. By the symmetry between
the segments around the inflection point, we demand that the $x$
value at point $V_1$ is twice that at the inflection point $I$.
Solving the resulting equation numerically gives the value for the
modulus as $k=1.1695$. The inflection point is identified by the
vanishing curvature of Eq.~\ref{eq:curvature} which corresponds to
a tangent angle of $\theta_I=2.052$.

The expression for the curvature (in Eq.~\ref{eq:curvature})
allows us to evaluate the bending energy of such a racquet head.
Since the bending energy is an integral over the squared
curvature, we can use Eq.~\ref{eq:curvature} to evaluate this
energy over any segment of the racquet head by integration. This
requires the numerical evaluation of an incomplete elliptic
integral of the second kind $E(\phi,k)$.
\begin{equation}
  \label{eq:bending}
  U = \frac{\beta\kappa}{k} \int d\theta\:\sqrt{1-k^2\sin^2(\theta/2)}
    = \frac{2\beta\kappa}{k}\:E(\theta/2,k)
\end{equation}
Adding up the symmetrical pieces of this solution for the entire
head yields the total bending energy of a complete racquet head
$U^{\text{head}} = A(\kappa/H)$ with $A$ representing the
numerical constant 18.3331. Thus, the bending energy in a racquet
head depends (apart from the value for the bending modulus
$\kappa$) only on its contour length. Note also that the bending
energy of the racquet head is very close to (but slightly below)
that of a circular ring with the same contour length $H$
($U^{\text{ring}} = 2\pi^2\kappa/H \approx 19.7392\;\kappa/H$).
Using circular rings as racquet heads would thus provide a
reasonable approximation for the calculation of conformational
energies, provided we neglect the penalty due to the sharp bends
at the neck.

The form of the solution in Eq.~\ref{eq:arclength} reveals that
our racquet head shape or elastica is unique, in the sense that it
is independent of any parameters in the problem. Both the
parametric head shape $s(\theta)$ and the bending energy
$U(\theta)$ are scaled by the factor $\beta$, related to the local
force balance at $V_2$. The overall size of the resulting shape is
merely scaled up or down, while its aspect ratio remains. Any
slender, uniform rod, subject to these boundary conditions, will
assume the described racquet head shape. We emphasize that the
size of the racquet head does \textit{not} depend on the overall
filament length $\lambda$, unlike in the case of the torus.

This dependence of the head size on the local force balance also
suggests that the following experiment should be possible, at
least in principle. Evaluating head sizes in a sample of partially
condensed filaments would measure the local interaction strength
between filaments, a quantity not easily found by other means.
This approach assumes of course, that the value for the bending
modulus (or equivalently the persistence length) is known from an
independent measurement.

\subsection{Bending Energy in Racquet Heads}

To evaluate the bending contribution to the conformational
energies we recall the expression for the bending energy in a head
of size $\chi$. Generalized to an $N$ bundle (which effectively
multiplies the bending modulus $\kappa$) the dimensionless bending
energy for an $N$ racquet head becomes $u^{\text{head}}_N(\chi) =
A(N/\chi)$ where $A$ is again the same numerical constant
evaluated previously from elliptic integrals. A stability analysis
and numerical minimizations found that ``perfect'' racquets (with
$\sigma=0$) are the solution for only a subset of all racquets.

%%%%%%%%%%%%%%%%%%%%%%%%%%%%%%%%%%%%%%%%%%%%%%%%%%%%%%%%%%%%%%%%%%%%
\begin{figure}[hbtp]
  \centering
  \includegraphics[width=0.5\columnwidth]{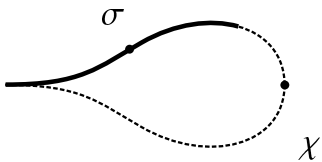}
  \caption{\label{fig:parthead}Partial overhang $\sigma$ (solid
  line) into an existing head of size $\chi$ (dashed). The ratio
  of $\sigma/\chi$ defines the fractional overhang $\varsigma$.
  This schematic shows the typical situation for racquet solutions
  with nonzero overhang, with the filament end located somewhere
  between the inflection and the halfway points (solid circles).}
\end{figure}
%%%%%%%%%%%%%%%%%%%%%%%%%%%%%%%%%%%%%%%%%%%%%%%%%%%%%%%%%%%%%%%%%%%%

In order to account for partial overhang into the heads, we need
to generalize the notion of the numerical prefactor $A$. This
``constant'' is really a function of the partial overhang. Due to
the scale invariance of our elastica, it is not surprising that
$A$ depends only on the \textit{relative} overhang $\varsigma
\equiv \sigma/\chi$. In terms of $\varsigma$ the four regions are
delimited by the following values: 0, 0.1627, 0.5, 0.8373, and
back to 1, measured from the neck. The three intermediate values
identify the two inflection points and the halfway point (the
origin in Fig.~\ref{fig:head}); note that these values are
measured in the opposite sense from the one defined in the figure.
Reconstruction of the piecewise solutions for any amount of
partial overhang $\sigma$ yields the expression
$u^{\text{partial}}(\sigma,\varsigma) = A(\varsigma)/\sigma$ with
the numerical prefactor $A \equiv A(\varsigma=1)$ generalized to
the \textit{function} $A(\varsigma)$.

\subsection{\label{subsec:evenodd}Even and Odd Racquets}

The surface energy terms for all racquets consist of several terms
with different coordination numbers in general. The only term the
even and odd cases share is the coordination in the neck, whose
length $\tau$ is shared by $(N+1)$ filaments. For the even
racquet, symmetry simplifies the expressions somewhat. In each of
its heads, we find a length $(\chi_n-\sigma_n)$ with coordination
$\alpha_n$ and the overhang piece $\sigma_n$ with coordination
$\alpha_{n+1}$ while the neck has the common coordination
$\alpha_{N+1}$ which leads to the full expression for the
conformational energy of the even racquet.
\begin{eqnarray}
  \label{eq:racquet_ue}
  u_N^{\text{even}} &=& A \left[\frac{N}{\chi_n}\right]
  + 2A(\sigma_n/\chi_n)\left[\frac{1}{\sigma_n}\right] \nonumber \\
  & & \mbox{} + 2\:\left[ \alpha_n(\chi_n - \sigma_n)
  + \alpha_{n+1}(\sigma_n) \right] \nonumber \\
  & & \mbox{} + \alpha_{N+1}(\tau)
\end{eqnarray}
The first two terms are the bending contributions for complete
heads and partial overhang, while the three following terms are
surface contributions for head segments and neck, respectively.
For the odd racquets, the expression becomes
\begin{eqnarray}
  \label{eq:racquet_uo}
  u_N^{\text{odd}} &=& A \left[\frac{p}{\chi_p} + \frac{q}{\chi_q}\right]
  + 2A(\sigma_p/\chi_p)\left[\frac{1}{\sigma_p}\right] \nonumber \\
  & & \mbox{} + \alpha_{p}(\chi_p - 2\sigma_p)
  + \alpha_{q}(\chi_q + 2\sigma_q) \nonumber \\
  & & \mbox{} + \alpha_{N+1} (\tau)\:.
\end{eqnarray}
In the even case, there are only two free variables, $\chi_n$ and
$\sigma_n$. We find their optimal values by simultaneous,
numerical minimization. In the odd case, the situation is
slightly different, since we lack the symmetry between heads.
However, we can make use of the fact that both filament ends (and
therefore any potential overhang $\sigma$) are on the left side.
This leaves the right head with a well-defined structure of $q$
filaments in the head and $N+1=2q$ filaments in the neck. Since
the head size is solely determined by the respective bundling
numbers in head and neck, we can determine right head sizes
independently of any overhang on the left by minimization in
terms of the various bundling and coordination numbers and $A$
only.
\begin{equation}
  \label{eq:righthead}
  \chi_q = \sqrt{\frac{2qA}{2\alpha_q - \alpha_{2q}}}
\end{equation}
We then numerically minimize over the two remaining free variables
$\chi_p$ and $\sigma_p$ of the left head. Plotting constant energy
contours as a function of the two free variables generally reveals
the approximate location of the relevant minimum, and their
coordinates were used as a starting point for the minimization
routine. This procedure finds two possible outcomes for both even
and odd cases. In the simpler case, the energy is minimized
without overhang ($\sigma=0$) and we recover the naively assumed,
perfect racquet structure. In the other case, we find a local
minimum with respect to $\sigma$ and $\chi$ for finite values of
overhang. In every case we have checked (up to $N=30$), these
solutions put the fractional overhang in the second region of the
racquet head, between the inflection and halfway points, as
indicated in Fig.~\ref{fig:parthead}.

%%%%%%%%%%%%%%%%%%%%%%%%%%%%%%%%%%%%%%%%%%%%%%%%%%%%%%%%%%%%%%%%%%%%
\begin{figure}[hbtp]
  \centering
  \includegraphics[width=0.5\columnwidth]{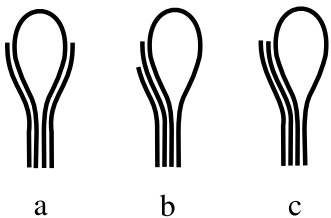}
  \caption{\label{fig:overhang}Three possible solutions for
  partial overhang into the left head of an odd racquet. Case (a)
  is the one described in the text. The more general case with
  different amounts of overhang on the same side (b) is always
  minimized by the arrangement in case (c) where the two ends
  coincide.}
\end{figure}
%%%%%%%%%%%%%%%%%%%%%%%%%%%%%%%%%%%%%%%%%%%%%%%%%%%%%%%%%%%%%%%%%%%%

Since both filament ends are on the same (left) side for the odd
racquets, there are several possible configurations for overhang
to be arranged, as shown in Fig.~\ref{fig:overhang}. The two
pieces of equal length $\sigma_p$ could be arranged symmetrically
on opposite sides of the head. Alternatively, the two pieces of
overhang can be on the same side, but not necessarily of equal
length. All the cases we examined are minimized for one unique
value of $\sigma$, corresponding to cases (a) and (c) in
Fig.~\ref{fig:overhang} which turn out to be degenerate in energy.
In retrospect we were thus justified to describe the odd racquet
with overhang generically as the symmetric case (a), while a
second (asymmetrical) solution, degenerate in overhang and energy,
exists.

\begin{figure}[htbp]
  \centering
  \includegraphics[width=\columnwidth]{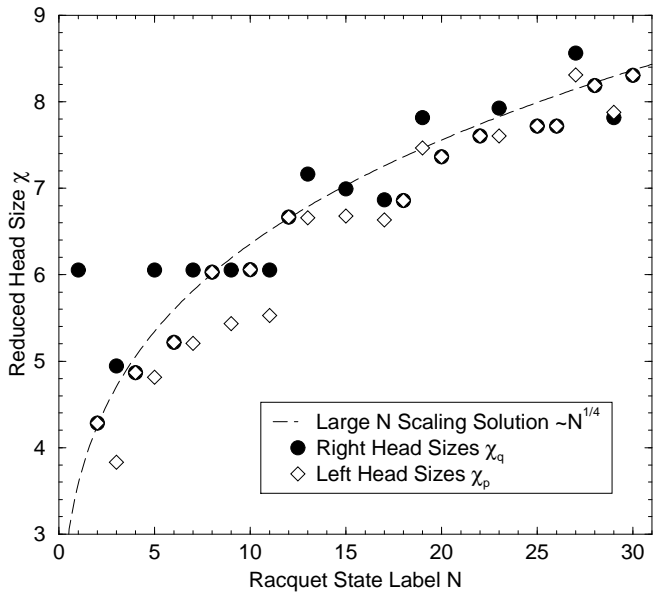}
  \caption{\label{fig:headsizes}Left and right head sizes versus
  the state label $N$. Pairs of even heads are of the same size,
  by symmetry. The general trend is for heads to grow with $N$, if
  not monotonically. Notice the convergence towards the asymptotic
  solution (dashed line, see Eqs.~\ref{eq:rac_prefactors}) with
  increasing $N$.}
\end{figure}

All racquet head sizes $\chi_p$ and $\chi_q$ found either by
direct calculation or by numerical minimization are displayed as a
function of the racquet state label $N$ in
Fig.~\ref{fig:headsizes} to show the general trend and their
convergence towards the large $N$ solution. Head sizes typically
increase with $N$, though not monotonically, and right heads are
typically larger than left for the odd racquets. Even racquet
heads are of the same size, by construction.

%%%%%%%%%%%%%%%%%%%%%%%%%%%%%%%%%%%%%%%%%%%%%%%%%%%%%%%%%%%%%%%%%%%%
%%%%%%%%%%%%%%%%%%%%%%%%%%%%%%%%%%%%%%%%%%%%%%%%%%%%%%%%%%%%%%%%%%%%

Since our minimization allowed only for extension into the heads
but not retraction of the filament ends back into the neck, we
tested the stability of racquets (up to $N=30$) to small
perturbations, subject to fixed overall length $\lambda$. We found
three types of results. In the simplest case, the racquets are
stable to any small change. These racquets (with
$N=1,2,3,11,15,17,20\ldots$) remain exact ($\sigma=0$). A second
class is identified by stability to retraction but not to
extension. These racquets (with $N=4-10,13,18,22\ldots$) develop
finite (positive) overhang. The remaining cases are the magic
numbers starting with 12 (namely $N=12,14,16,19,21\ldots$) which
are unstable (or marginally stable) to retraction along the neck.
A subset of these states are unstable to extension, and solutions
with finite, positive overhang exist. However, those cases which
are stable to extension have no metastable solution at all. We
thus conclude that no racquet solutions exist for
$N=14,16,21,24\ldots$ and these states are omitted from our energy
spectra (Fig.~\ref{fig:rac_scaling}) and the series of head sizes
(Fig.~\ref{fig:headsizes}).

\subsection{\label{subsec:racquet_scaling}Large $N$ Limit and Scaling}

\begin{figure}[hbtp]
  \centering
  \includegraphics[width=0.8\columnwidth]{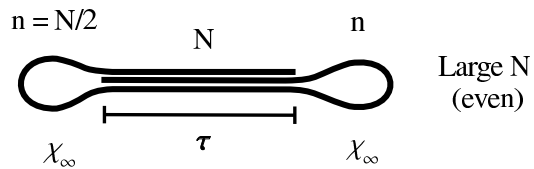}
  \caption{\label{fig:largeN}Schematic racquet in the limit of
  large $N$ where we assume the symmetry of the even racquet and
  neglect any extra overhang.}
\end{figure}

We perform the analogous calculation to that done for the torus
states in Section~\ref{subsec:torus_scaling}, under the assumption
that bundles form hexagonal cross sections as their bundling
numbers $N$ become large, to find the behavior of the racquet
energies in the same limit. The result is shown as the dashed
lines in Figs.~\ref{fig:rac_scaling} and~\ref{fig:headsizes}. To
compute it, we assume that the large $N$ racquet be even and
without overhang ($\sigma=0$) as differences between bundles of
nearly the same number of filaments vanish in this limit. This
even racquet has a neck length $\tau$ and a limiting head size
$\chi_\infty$ for large bundling numbers $n$ in the heads and $N$
in the neck (see Fig~\ref{fig:largeN}).

%%%%%%%%%%%%%%%%%%%%%%%%%%%%%%%%%%%%%%%%%%%%%%%%%%%%%%%%%%%%%%%%%%%%
%%%%%%%%%%%%%%%%%%%%%%%%%%%%%%%%%%%%%%%%%%%%%%%%%%%%%%%%%%%%%%%%%%%%

%%%%%%%%%%%%%%%%%%%%%%%%%%%%%%%%%%%%%%%%%%%%%%%%%%%%%%%%%%%%%%%%%%%%
\begin{figure*}[hbtp]
  \centering
  \includegraphics[width=0.9\textwidth]{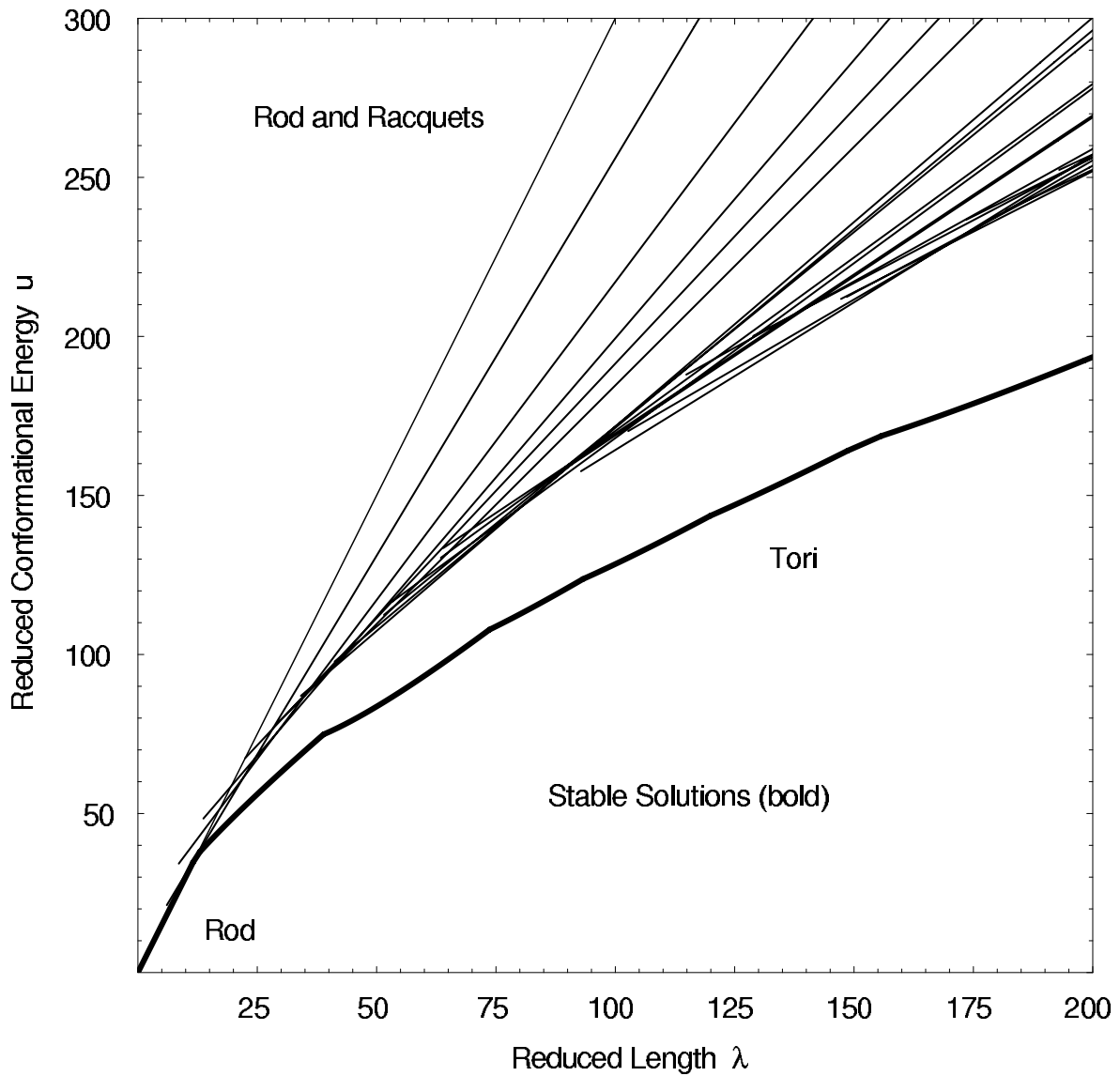}
  \caption{\label{fig:cascade}Spectrum of rod, racquet and torus
  states shown as conformational energy versus filament length in
  reduced units. Only the rod (at small $\lambda$) and the tori
  (for all $\lambda$ beyond a transition point) are globally
  stable states. Notice the rather large gap between the spectrum
  of racquet states and the stable torus solution. Metastable
  torus solutions are omitted for clarity.}
\end{figure*}
%%%%%%%%%%%%%%%%%%%%%%%%%%%%%%%%%%%%%%%%%%%%%%%%%%%%%%%%%%%%%%%%%%%%

Since the size of the heads depends only on the balance of forces
at the point where the head and neck bundles meet, we can
calculate the optimal head size $\chi_{\infty}$ as in
Eq.~\ref{eq:righthead} for the right head of an odd racquet. As in
Section~\ref{subsec:torus_scaling}, we determine the optimal
bundling number $N_{\text{opt}}(\lambda)$ by minimizing the energy
with respect to $N$, which yields the scaling results with
prefactors as functions of $\lambda$ only.
\begin{subequations}
  \label{eq:rac_prefactors}
  \begin{eqnarray}
    N_{\text{opt}}  &\approx& 0.303\:\lambda^{4/5} \\
    \chi_\infty     &\approx& 2.653\:\lambda^{1/5} \label{eq:chinfty_pref} \\
    u_\infty        &\approx& 10.482\:\lambda^{3/5}
  \end{eqnarray}
\end{subequations}

Knowing the head size $\chi_\infty$, we can calculate the lower
limit of validity $\lambda_{\text{low}}$ in a scaling sense. This
allows us to compare the expressions for the filament length from
minimization ($\lambda_{\text{opt}} \approx 4.442\:N^{5/4}$) with
the length found by simply removing the neck altogether
($\lambda_{\text{low}} \approx 3.575\:N^{5/4}$). Since the optimal
length $\lambda_{\text{opt}}$ exceeds the minimal length
$\lambda_{\text{low}}$, a large $N$ racquet will be one with a
finite neck. This is an important result since it hints at the
evolution of very long chains as they condense into racquets with
increasingly larger $N$. In fact, we can estimate the growth of
the neck length $\tau_\infty$ from the difference between the
prefactors in $\lambda_{\text{low}}$ and $\lambda_{\text{opt}}$.
Its scaling is given by $\tau_{\infty} \approx
0.644\:\lambda^{1/5}$. Thus, the neck grows with the same power of
$\lambda$ as the heads but with a smaller prefactor. We may have
anticipated that the growing heads provide a simple pathway
towards the torus, as the inevitable limit of the heads growing at
the expense of the neck. For a fixed filament length, the neck
would have had to shrink to zero with increasing $N$, opening the
structure up to form a torus. For the particular racquet solutions
shown in Fig.~\ref{fig:rac_scaling} we notice that the end points
are relatively dense and represent, at times, the lowest point in
the spectrum of states. Especially for such states, it is still
true that their neck can shrink to very small or vanishing
lengths, depending on $\lambda$. Thermal fluctuations can then
lead to the opening up of the neck to form a torus. Yet, even if
the limit of large $N$ does not provide an absolutely compelling
pathway for the collapse to the torus, we now appreciate the
energetics involved.

\section{\label{sec:racqvstori}Discussion: Racquets versus Tori}

Fig.~\ref{fig:cascade} shows the individual racquet solutions of
Fig.~\ref{fig:rac_scaling} now compared to the stable torus ground
states found in Section~\ref{subsec:stability}. In anticipation of
these results, we described the lowest metastable torus state over
any range of $\lambda$ as the ground state of the system.
Fig.~\ref{fig:cascade} confirms this claim by direct comparison of
racquets and tori. In addition, we found that the large $N$
solutions for tori and racquets both grow as $\lambda^{3/5}$ but
with different prefactors. In combination with the close agreement
between particular solutions and the large $N$ limit, this
strongly suggests that the torus remains the ground state for all
$\lambda$ beyond the transition point ($\lambda=11.543$). Only for
shorter chains is the rod the ground state.

%%%%%%%%%%%%%%%%%%%%%%%%%%%%%%%%%%%%%%%%%%%%%%%%%%%%%%%%%%%%%%%%%%%%
\begin{figure}[hbtp]
  \centering
  \includegraphics[width=\columnwidth]{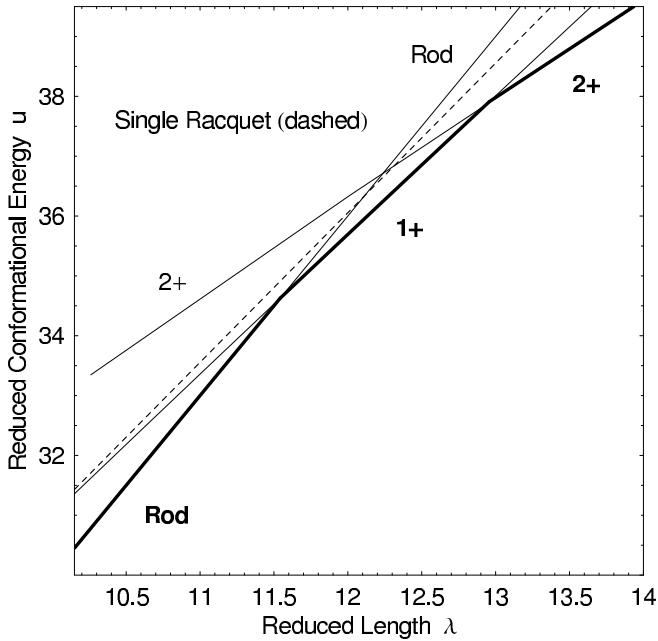}
  \caption{\label{fig:detail}Close-up of the rod, racquet and
  torus solutions in the region where they are closest to each
  other. Racquets are indeed never stable, though their energy is
  very close to both the rod and the tori in this region.}
\end{figure}
%%%%%%%%%%%%%%%%%%%%%%%%%%%%%%%%%%%%%%%%%%%%%%%%%%%%%%%%%%%%%%%%%%%%

There appears to be only one region where the energies of
racquets and tori are even close, at the very low values of
$\lambda$ near the transition point. Fig.~\ref{fig:detail} shows
the relevant region in detail. The $N=1$ racquet solution comes
extremely close to the solutions for both the rod ($N=0$) as well
as the 1+ torus, but remains above. Thus the only stable (ground
state) solutions for this system (in the absence of thermal
fluctuations) are the rod at small $\lambda$ and the tori
everywhere beyond the transition point. At energies above this
ground state, we see a dense spectrum of metastable solutions,
made up of other (metastable) torus (see Fig.~\ref{fig:toroids}
for details) and increasing numbers of racquet states.

For fixed conditions we need only consider a vertical slice
through the spectrum of energies. Along such a line, we can
imagine a filament cascading down from an extended, rod-like
configuration, through various metastable intermediates, while
lowering its energy along the way. Our calculations do not of
course capture the entire physical picture, as we neglect filament
size in the bundling and our states are calculated in the absence
of thermal undulations. So far we have no estimate of the energy
barriers between the metastable intermediates. However, the
dynamical simulation results~\cite{schnurr2000epl} suggest that
these barriers as well as the energy gaps between states are large
compared to $k_BT$: transitions that increase $N$ are infrequent
and sharp, while transitions in the opposite direction are
essentially never observed. This is especially true for the
transition from the racquet spectrum to a torus, indicating that
this energy gap is even larger for the parameters chosen in the
simulation. This picture is consistent with the analytic results
in Fig.~\ref{fig:cascade} which clearly shows the large gap
stabilizing the ground state. The results of our analysis thus
nicely corroborate, at least qualitatively, the results of our
prior computer simulations as well as their relevance to the
condensation of stiff chains.

We would like to note that the shape of condensed filaments may
depend on the nature and molecular structure of the condensing
agent. Our study only addresses an interaction that is uniform
along the filament, such as the effect due to a poor solvent.
Other systems, with more point-like organizing centers, have been
shown to exhibit intricate multi-leaf or flower
patterns~\cite{fang1998mica, schiessel2000}.

Our observations suggest that the pathway for the collapse of
extended chains into condensed structures via intermediate racquet
states is a viable, even generic alternative to the perhaps more
immediately guessed direct winding up upon the meeting of filament
ends at an obtuse angle. Some of the simulations show this latter
collapse pathway, but it is much less frequent. Furthermore, this
cascade picture through which our calculations reinforce and at
least partially explain the simulation results, seems robust. We
find this cascade through intermediate states even for a much more
naive treatment of the poor solvent interaction used in a first
pass. The individual curves (\textit{e.g.} in
Fig.~\ref{fig:cascade}) are shifted but show a qualitatively
similar picture. The generic cascade through metastable
intermediates is so dominant to be retained regardless of the
detailed realization of the interactions.

\begin{acknowledgments}
The authors wish to thank David Williams for helpful discussions,
including sharing aspects of related work~\cite{pereira2000}. This
work was supported in part by the Whitaker Foundation and by NSF
Grant Nos.\ DMR-9257544, INT-9605179. BS acknowledges support from
the Minerva Foundation (Max Planck Society).
\end{acknowledgments}


\begin{thebibliography}{47}
\expandafter\ifx\csname
natexlab\endcsname\relax\def\natexlab#1{#1}\fi
\expandafter\ifx\csname bibnamefont\endcsname\relax
  \def\bibnamefont#1{#1}\fi
\expandafter\ifx\csname bibfnamefont\endcsname\relax
  \def\bibfnamefont#1{#1}\fi
\expandafter\ifx\csname citenamefont\endcsname\relax
  \def\citenamefont#1{#1}\fi
\expandafter\ifx\csname url\endcsname\relax
  \def\url#1{\texttt{#1}}\fi
\expandafter\ifx\csname
urlprefix\endcsname\relax\def\urlprefix{URL }\fi
\providecommand{\bibinfo}[2]{#2}
\providecommand{\eprint}[2][]{\url{#2}}

\bibitem[{\citenamefont{deGennes}(1979)}]{degennes1979scaling}
\bibinfo{author}{\bibfnamefont{P.-G.} \bibnamefont{deGennes}},
  \emph{\bibinfo{title}{Scaling Concepts in Polymer Physics}}
  (\bibinfo{publisher}{Cornell University Press}, \bibinfo{address}{Ithaca},
  \bibinfo{year}{1979}).

\bibitem[{\citenamefont{Doi and Edwards}(1988)}]{doiedwards1988}
\bibinfo{author}{\bibfnamefont{M.}~\bibnamefont{Doi}} \bibnamefont{and}
  \bibinfo{author}{\bibfnamefont{S.~F.} \bibnamefont{Edwards}},
  \emph{\bibinfo{title}{The Theory of Polymer Dynamics}}
  (\bibinfo{publisher}{Clarendon Press}, \bibinfo{address}{Oxford},
  \bibinfo{year}{1988}).

\bibitem[{\citenamefont{Grosberg and Khokhlov}(1994)}]{grosberg1994statphys}
\bibinfo{author}{\bibfnamefont{A.~Y.} \bibnamefont{Grosberg}} \bibnamefont{and}
  \bibinfo{author}{\bibfnamefont{A.~R.} \bibnamefont{Khokhlov}},
  \emph{\bibinfo{title}{Statistical Physics of Macromolecules}}, AIP Series in
  Polymers and Complex Materials (\bibinfo{publisher}{AIP Press},
  \bibinfo{address}{New York}, \bibinfo{year}{1994}).

\bibitem[{\citenamefont{deGennes}(1985)}]{degennes1985kin}
\bibinfo{author}{\bibfnamefont{P.~G.} \bibnamefont{deGennes}},
  \bibinfo{journal}{J Phys Lett - Paris}
  \textbf{\bibinfo{volume}{\textbf{46}}}, \bibinfo{pages}{L639}
  (\bibinfo{year}{1985}).

\bibitem[{\citenamefont{Ostrovsky and Bar-Yam}(1994)}]{ostrovsky1994}
\bibinfo{author}{\bibfnamefont{B.}~\bibnamefont{Ostrovsky}} \bibnamefont{and}
  \bibinfo{author}{\bibfnamefont{Y.}~\bibnamefont{Bar-Yam}},
  \bibinfo{journal}{Europhys. Lett.} \textbf{\bibinfo{volume}{\textbf{25}}},
  \bibinfo{pages}{409} (\bibinfo{year}{1994}).

\bibitem[{\citenamefont{Buguin et~al.}(1996)\citenamefont{Buguin,
  Brochard-Wyart, and deGennes}}]{buguin1996}
\bibinfo{author}{\bibfnamefont{A.}~\bibnamefont{Buguin}},
  \bibinfo{author}{\bibfnamefont{F.}~\bibnamefont{Brochard-Wyart}},
  \bibnamefont{and} \bibinfo{author}{\bibfnamefont{P.~G.}
  \bibnamefont{deGennes}}, \bibinfo{journal}{Comptes Rendus de l'Acad\'emie des
  Sciences S\'erie II Fascicule B - M\'ecanique Physique Chimie Astronomie}
  \textbf{\bibinfo{volume}{\textbf{322}}}, \bibinfo{pages}{741}
  (\bibinfo{year}{1996}).

\bibitem[{\citenamefont{Dawson et~al.}(1997)\citenamefont{Dawson, Timoshenko,
  and Kuznetsov}}]{dawson1997}
\bibinfo{author}{\bibfnamefont{K.~A.} \bibnamefont{Dawson}},
  \bibinfo{author}{\bibfnamefont{E.~G.} \bibnamefont{Timoshenko}},
  \bibnamefont{and} \bibinfo{author}{\bibfnamefont{Y.~A.}
  \bibnamefont{Kuznetsov}}, \bibinfo{journal}{Physica A}
  \textbf{\bibinfo{volume}{\textbf{236}}}, \bibinfo{pages}{58}
  (\bibinfo{year}{1997}).

\bibitem[{\citenamefont{Halperin and Goldbart}(2000)}]{halperin2000}
\bibinfo{author}{\bibfnamefont{A.}~\bibnamefont{Halperin}} \bibnamefont{and}
  \bibinfo{author}{\bibfnamefont{P.~M.} \bibnamefont{Goldbart}},
  \bibinfo{journal}{Phys. Rev. E} \textbf{\bibinfo{volume}{\textbf{61}}},
  \bibinfo{pages}{565} (\bibinfo{year}{2000}).

\bibitem[{\citenamefont{Lee and Thirumalai}(2001)}]{lee2001}
\bibinfo{author}{\bibfnamefont{N.}~\bibnamefont{Lee}} \bibnamefont{and}
  \bibinfo{author}{\bibfnamefont{D.}~\bibnamefont{Thirumalai}},
  \bibinfo{journal}{Macromolecules} \textbf{\bibinfo{volume}{\textbf{34}}},
  \bibinfo{pages}{3446} (\bibinfo{year}{2001}).

\bibitem[{\citenamefont{Chu et~al.}(1995)\citenamefont{Chu, Ying, and
  Grosberg}}]{chu1995}
\bibinfo{author}{\bibfnamefont{B.}~\bibnamefont{Chu}},
  \bibinfo{author}{\bibfnamefont{Q.~C.} \bibnamefont{Ying}}, \bibnamefont{and}
  \bibinfo{author}{\bibfnamefont{A.~Y.} \bibnamefont{Grosberg}},
  \bibinfo{journal}{Macromolecules} \textbf{\bibinfo{volume}{\textbf{28}}},
  \bibinfo{pages}{180} (\bibinfo{year}{1995}).

\bibitem[{\citenamefont{Abrams et~al.}(2001)\citenamefont{Abrams, Lee, and
  Obukhov}}]{abrams2001}
\bibinfo{author}{\bibfnamefont{C.~F.} \bibnamefont{Abrams}},
  \bibinfo{author}{\bibfnamefont{N.}~\bibnamefont{Lee}}, \bibnamefont{and}
  \bibinfo{author}{\bibfnamefont{S.}~\bibnamefont{Obukhov}}
  (\bibinfo{year}{2001}), \eprint{cond-mat/0110491}.

\bibitem[{\citenamefont{Grosberg}(1979)}]{grosberg1979}
\bibinfo{author}{\bibfnamefont{A.~Y.} \bibnamefont{Grosberg}},
  \bibinfo{journal}{Biofizika} \textbf{\bibinfo{volume}{\textbf{24}}},
  \bibinfo{pages}{32} (\bibinfo{year}{1979}).

\bibitem[{\citenamefont{Bloomfield}(1991)}]{bloomfield1991}
\bibinfo{author}{\bibfnamefont{V.~A.} \bibnamefont{Bloomfield}},
  \bibinfo{journal}{Biopolymers} \textbf{\bibinfo{volume}{\textbf{31}}},
  \bibinfo{pages}{1471} (\bibinfo{year}{1991}).

\bibitem[{\citenamefont{Bloomfield}(1997)}]{bloomfield1997}
\bibinfo{author}{\bibfnamefont{V.~A.} \bibnamefont{Bloomfield}},
  \bibinfo{journal}{Biopolymers} \textbf{\bibinfo{volume}{\textbf{44}}},
  \bibinfo{pages}{269} (\bibinfo{year}{1997}).

\bibitem[{\citenamefont{Vasilevskaya et~al.}(1997)\citenamefont{Vasilevskaya,
  Khokhlov, Kidoaki, and Yoshikawa}}]{vasil1997}
\bibinfo{author}{\bibfnamefont{V.~V.} \bibnamefont{Vasilevskaya}},
  \bibinfo{author}{\bibfnamefont{A.~R.} \bibnamefont{Khokhlov}},
  \bibinfo{author}{\bibfnamefont{S.}~\bibnamefont{Kidoaki}}, \bibnamefont{and}
  \bibinfo{author}{\bibfnamefont{K.}~\bibnamefont{Yoshikawa}},
  \bibinfo{journal}{Biopolymers} \textbf{\bibinfo{volume}{\textbf{41}}},
  \bibinfo{pages}{51} (\bibinfo{year}{1997}).

\bibitem[{\citenamefont{Li et~al.}(1992)\citenamefont{Li, Fan, and
  Ding}}]{lifanding1992}
\bibinfo{author}{\bibfnamefont{A.~Z.} \bibnamefont{Li}},
  \bibinfo{author}{\bibfnamefont{T.~Y.} \bibnamefont{Fan}}, \bibnamefont{and}
  \bibinfo{author}{\bibfnamefont{M.}~\bibnamefont{Ding}},
  \bibinfo{journal}{Sci. China Ser. B-Chem.}
  \textbf{\bibinfo{volume}{\textbf{35}}}, \bibinfo{pages}{169}
  (\bibinfo{year}{1992}).

\bibitem[{\citenamefont{Fang and Hoh}(1998{\natexlab{a}})}]{fang1998surf}
\bibinfo{author}{\bibfnamefont{Y.}~\bibnamefont{Fang}} \bibnamefont{and}
  \bibinfo{author}{\bibfnamefont{J.~H.} \bibnamefont{Hoh}},
  \bibinfo{journal}{Nucleic Acids Res.} \textbf{\bibinfo{volume}{\textbf{26}}},
  \bibinfo{pages}{588} (\bibinfo{year}{1998}{\natexlab{a}}).

\bibitem[{\citenamefont{Shen et~al.}(2000)\citenamefont{Shen, Downing, Balhorn,
  and Hud}}]{shen2000}
\bibinfo{author}{\bibfnamefont{M.~R.} \bibnamefont{Shen}},
  \bibinfo{author}{\bibfnamefont{K.~H.} \bibnamefont{Downing}},
  \bibinfo{author}{\bibfnamefont{R.}~\bibnamefont{Balhorn}}, \bibnamefont{and}
  \bibinfo{author}{\bibfnamefont{N.~V.} \bibnamefont{Hud}},
  \bibinfo{journal}{J. Am. Chem. Soc.} \textbf{\bibinfo{volume}{\textbf{122}}},
  \bibinfo{pages}{4833} (\bibinfo{year}{2000}).

\bibitem[{\citenamefont{Martin et~al.}(2000)\citenamefont{Martin, Davies,
  Rackstraw, Roberts, Stolnik, Tendler, and Williams}}]{martin2000}
\bibinfo{author}{\bibfnamefont{A.~L.} \bibnamefont{Martin}},
  \bibinfo{author}{\bibfnamefont{M.~C.} \bibnamefont{Davies}},
  \bibinfo{author}{\bibfnamefont{B.~J.} \bibnamefont{Rackstraw}},
  \bibinfo{author}{\bibfnamefont{C.~J.} \bibnamefont{Roberts}},
  \bibinfo{author}{\bibfnamefont{S.}~\bibnamefont{Stolnik}},
  \bibinfo{author}{\bibfnamefont{S.~J.~B.} \bibnamefont{Tendler}},
  \bibnamefont{and} \bibinfo{author}{\bibfnamefont{P.~M.}
  \bibnamefont{Williams}}, \bibinfo{journal}{FEBS Lett.}
  \textbf{\bibinfo{volume}{\textbf{480}}}, \bibinfo{pages}{106}
  (\bibinfo{year}{2000}).

\bibitem[{\citenamefont{Liu et~al.}(2001)\citenamefont{Liu, Wang, Lin, Li, Xu,
  Wei, Wang, and Bai}}]{liu2001interm}
\bibinfo{author}{\bibfnamefont{D.}~\bibnamefont{Liu}},
  \bibinfo{author}{\bibfnamefont{C.}~\bibnamefont{Wang}},
  \bibinfo{author}{\bibfnamefont{Z.}~\bibnamefont{Lin}},
  \bibinfo{author}{\bibfnamefont{J.~W.} \bibnamefont{Li}},
  \bibinfo{author}{\bibfnamefont{B.}~\bibnamefont{Xu}},
  \bibinfo{author}{\bibfnamefont{Z.~Q.} \bibnamefont{Wei}},
  \bibinfo{author}{\bibfnamefont{Z.~G.} \bibnamefont{Wang}}, \bibnamefont{and}
  \bibinfo{author}{\bibfnamefont{C.~L.} \bibnamefont{Bai}},
  \bibinfo{journal}{Surf. Interface Anal.}
  \textbf{\bibinfo{volume}{\textbf{32}}}, \bibinfo{pages}{15}
  (\bibinfo{year}{2001}).

\bibitem[{\citenamefont{Hansma}(2001)}]{hansma2001surf}
\bibinfo{author}{\bibfnamefont{H.~G.} \bibnamefont{Hansma}},
  \bibinfo{journal}{Annu. Rev. Phys. Chem.}
  \textbf{\bibinfo{volume}{\textbf{52}}}, \bibinfo{pages}{71}
  (\bibinfo{year}{2001}).

\bibitem[{\citenamefont{Noguchi et~al.}(1996)\citenamefont{Noguchi, Saito,
  Kidoaki, and Yoshikawa}}]{noguchi1996}
\bibinfo{author}{\bibfnamefont{H.}~\bibnamefont{Noguchi}},
  \bibinfo{author}{\bibfnamefont{S.}~\bibnamefont{Saito}},
  \bibinfo{author}{\bibfnamefont{S.}~\bibnamefont{Kidoaki}}, \bibnamefont{and}
  \bibinfo{author}{\bibfnamefont{K.}~\bibnamefont{Yoshikawa}},
  \bibinfo{journal}{Chem. Phys. Lett.} \textbf{\bibinfo{volume}{\textbf{261}}},
  \bibinfo{pages}{527} (\bibinfo{year}{1996}).

\bibitem[{\citenamefont{Byrne et~al.}(1998)\citenamefont{Byrne, Timoshenko, and
  Dawson}}]{byrne1998kin}
\bibinfo{author}{\bibfnamefont{A.}~\bibnamefont{Byrne}},
  \bibinfo{author}{\bibfnamefont{E.~G.} \bibnamefont{Timoshenko}},
  \bibnamefont{and} \bibinfo{author}{\bibfnamefont{K.~A.}
  \bibnamefont{Dawson}}, \bibinfo{journal}{Nuovo Cimento della Societa Italiana
  di Fisica D} \textbf{\bibinfo{volume}{\textbf{20}}}, \bibinfo{pages}{2289}
  (\bibinfo{year}{1998}).

\bibitem[{\citenamefont{Noguchi and Yoshikawa}(2000)}]{noguchi2000bd}
\bibinfo{author}{\bibfnamefont{H.}~\bibnamefont{Noguchi}} \bibnamefont{and}
  \bibinfo{author}{\bibfnamefont{K.}~\bibnamefont{Yoshikawa}},
  \bibinfo{journal}{J. Chem. Phys.} \textbf{\bibinfo{volume}{\textbf{113}}},
  \bibinfo{pages}{854} (\bibinfo{year}{2000}).

\bibitem[{\citenamefont{Stevens}(2001)}]{stevens2001}
\bibinfo{author}{\bibfnamefont{M.~J.} \bibnamefont{Stevens}},
  \bibinfo{journal}{Biophys. J.} \textbf{\bibinfo{volume}{\textbf{80}}},
  \bibinfo{pages}{130} (\bibinfo{year}{2001}).

\bibitem[{\citenamefont{Hud}(1995)}]{hud1995pha}
\bibinfo{author}{\bibfnamefont{N.~V.} \bibnamefont{Hud}},
  \bibinfo{journal}{Biophys. J.} \textbf{\bibinfo{volume}{\textbf{69}}},
  \bibinfo{pages}{1355} (\bibinfo{year}{1995}).

\bibitem[{\citenamefont{Hud et~al.}(1995)\citenamefont{Hud, Downing, and
  Balhorn}}]{hud1995rad}
\bibinfo{author}{\bibfnamefont{N.~V.} \bibnamefont{Hud}},
  \bibinfo{author}{\bibfnamefont{K.~H.} \bibnamefont{Downing}},
  \bibnamefont{and} \bibinfo{author}{\bibfnamefont{R.}~\bibnamefont{Balhorn}},
  \bibinfo{journal}{Proc. Natl. Acad. Sci. U. S. A.}
  \textbf{\bibinfo{volume}{\textbf{92}}}, \bibinfo{pages}{3581}
  (\bibinfo{year}{1995}).

\bibitem[{\citenamefont{Ubbink and Odijk}(1995)}]{ubbink1995}
\bibinfo{author}{\bibfnamefont{J.}~\bibnamefont{Ubbink}} \bibnamefont{and}
  \bibinfo{author}{\bibfnamefont{T.}~\bibnamefont{Odijk}},
  \bibinfo{journal}{Biophys. J.} \textbf{\bibinfo{volume}{\textbf{68}}},
  \bibinfo{pages}{54} (\bibinfo{year}{1995}).

\bibitem[{\citenamefont{Ubbink and Odijk}(1996)}]{ubbink1996def}
\bibinfo{author}{\bibfnamefont{J.}~\bibnamefont{Ubbink}} \bibnamefont{and}
  \bibinfo{author}{\bibfnamefont{T.}~\bibnamefont{Odijk}},
  \bibinfo{journal}{Europhys. Lett.} \textbf{\bibinfo{volume}{\textbf{33}}},
  \bibinfo{pages}{353} (\bibinfo{year}{1996}).

\bibitem[{\citenamefont{Park et~al.}(1998)\citenamefont{Park, Harries, and
  Gelbart}}]{park1998}
\bibinfo{author}{\bibfnamefont{S.~Y.} \bibnamefont{Park}},
  \bibinfo{author}{\bibfnamefont{D.}~\bibnamefont{Harries}}, \bibnamefont{and}
  \bibinfo{author}{\bibfnamefont{W.~M.} \bibnamefont{Gelbart}},
  \bibinfo{journal}{Biophys. J.} \textbf{\bibinfo{volume}{\textbf{75}}},
  \bibinfo{pages}{714} (\bibinfo{year}{1998}).

\bibitem[{\citenamefont{Golo et~al.}(1998)\citenamefont{Golo, Kats, and
  Yevdokimov}}]{golo1998}
\bibinfo{author}{\bibfnamefont{V.~L.} \bibnamefont{Golo}},
  \bibinfo{author}{\bibfnamefont{E.~I.} \bibnamefont{Kats}}, \bibnamefont{and}
  \bibinfo{author}{\bibfnamefont{Y.~M.} \bibnamefont{Yevdokimov}},
  \bibinfo{journal}{Journal of Biomolecular Structure \& Dynamics}
  \textbf{\bibinfo{volume}{\textbf{15}}}, \bibinfo{pages}{757}
  (\bibinfo{year}{1998}).

\bibitem[{\citenamefont{Fang and Hoh}(1998{\natexlab{b}})}]{fang1998mica}
\bibinfo{author}{\bibfnamefont{Y.}~\bibnamefont{Fang}} \bibnamefont{and}
  \bibinfo{author}{\bibfnamefont{J.~H.} \bibnamefont{Hoh}},
  \bibinfo{journal}{J. Am. Chem. Soc.} \textbf{\bibinfo{volume}{\textbf{120}}},
  \bibinfo{pages}{8903} (\bibinfo{year}{1998}{\natexlab{b}}).

\bibitem[{\citenamefont{He et~al.}(2000)\citenamefont{He, Arscott, and
  Bloomfield}}]{he2000}
\bibinfo{author}{\bibfnamefont{S.~Q.} \bibnamefont{He}},
  \bibinfo{author}{\bibfnamefont{P.~G.} \bibnamefont{Arscott}},
  \bibnamefont{and} \bibinfo{author}{\bibfnamefont{V.~A.}
  \bibnamefont{Bloomfield}}, \bibinfo{journal}{Biopolymers}
  \textbf{\bibinfo{volume}{\textbf{53}}}, \bibinfo{pages}{329}
  (\bibinfo{year}{2000}).

\bibitem[{\citenamefont{Schnurr et~al.}(2000)\citenamefont{Schnurr, MacKintosh,
  and Williams}}]{schnurr2000epl}
\bibinfo{author}{\bibfnamefont{B.}~\bibnamefont{Schnurr}},
  \bibinfo{author}{\bibfnamefont{F.~C.} \bibnamefont{MacKintosh}},
  \bibnamefont{and} \bibinfo{author}{\bibfnamefont{D.~R.~M.}
  \bibnamefont{Williams}}, \bibinfo{journal}{Europhys. Lett.}
  \textbf{\bibinfo{volume}{\textbf{51}}}, \bibinfo{pages}{279}
  (\bibinfo{year}{2000}).

\bibitem[{\citenamefont{Schnurr}(2000)}]{schnurrthesis}
\bibinfo{author}{\bibfnamefont{B.}~\bibnamefont{Schnurr}}, Ph.D. thesis,
  \bibinfo{school}{University of Michigan} (\bibinfo{year}{2000}).

\bibitem[{\citenamefont{Evdokimov et~al.}(1972)\citenamefont{Evdokimov,
  Platonov, Varshavsky, and Tikhonenko}}]{evdokimov1972}
\bibinfo{author}{\bibfnamefont{Y.~M.} \bibnamefont{Evdokimov}},
  \bibinfo{author}{\bibfnamefont{A.~L.} \bibnamefont{Platonov}},
  \bibinfo{author}{\bibfnamefont{Y.~M.} \bibnamefont{Varshavsky}},
  \bibnamefont{and} \bibinfo{author}{\bibfnamefont{A.~S.}
  \bibnamefont{Tikhonenko}}, \bibinfo{journal}{FEBS Lett.}
  \textbf{\bibinfo{volume}{\textbf{23}}}, \bibinfo{pages}{180}
  (\bibinfo{year}{1972}).

\bibitem[{\citenamefont{Maniatis et~al.}(1974)\citenamefont{Maniatis, Venable,
  and Lerman}}]{maniatis1974}
\bibinfo{author}{\bibfnamefont{T.}~\bibnamefont{Maniatis}},
  \bibinfo{author}{\bibfnamefont{J.~H.} \bibnamefont{Venable}},
  \bibnamefont{and} \bibinfo{author}{\bibfnamefont{L.~S.}
  \bibnamefont{Lerman}}, \bibinfo{journal}{J. Mol. Biol.}
  \textbf{\bibinfo{volume}{\textbf{84}}}, \bibinfo{pages}{37}
  (\bibinfo{year}{1974}).

\bibitem[{\citenamefont{Evdokimov et~al.}(1976)\citenamefont{Evdokimov,
  Pyatigorskaya, Polyvtsev, Akimenko, Kadykov, Tsvankin, and
  Varshavsky}}]{evdokimov1976}
\bibinfo{author}{\bibfnamefont{Y.~M.} \bibnamefont{Evdokimov}},
  \bibinfo{author}{\bibfnamefont{T.~L.} \bibnamefont{Pyatigorskaya}},
  \bibinfo{author}{\bibfnamefont{O.~F.} \bibnamefont{Polyvtsev}},
  \bibinfo{author}{\bibfnamefont{N.~M.} \bibnamefont{Akimenko}},
  \bibinfo{author}{\bibfnamefont{V.~A.} \bibnamefont{Kadykov}},
  \bibinfo{author}{\bibfnamefont{D.~Y.} \bibnamefont{Tsvankin}},
  \bibnamefont{and} \bibinfo{author}{\bibfnamefont{Y.~M.}
  \bibnamefont{Varshavsky}}, \bibinfo{journal}{Nucleic Acids Res.}
  \textbf{\bibinfo{volume}{\textbf{3}}}, \bibinfo{pages}{2353}
  (\bibinfo{year}{1976}).

\bibitem[{\citenamefont{Podgornik et~al.}(1989)\citenamefont{Podgornik, Rau,
  and Parsegian}}]{podgornik1989}
\bibinfo{author}{\bibfnamefont{R.}~\bibnamefont{Podgornik}},
  \bibinfo{author}{\bibfnamefont{D.~C.} \bibnamefont{Rau}}, \bibnamefont{and}
  \bibinfo{author}{\bibfnamefont{V.~A.} \bibnamefont{Parsegian}},
  \bibinfo{journal}{Macromolecules} \textbf{\bibinfo{volume}{\textbf{22}}},
  \bibinfo{pages}{1780} (\bibinfo{year}{1989}).

\bibitem[{\citenamefont{Podgornik and Parsegian}(1990)}]{podgornik1990}
\bibinfo{author}{\bibfnamefont{R.}~\bibnamefont{Podgornik}} \bibnamefont{and}
  \bibinfo{author}{\bibfnamefont{V.~A.} \bibnamefont{Parsegian}},
  \bibinfo{journal}{Macromolecules} \textbf{\bibinfo{volume}{\textbf{23}}},
  \bibinfo{pages}{2265} (\bibinfo{year}{1990}).

\bibitem[{\citenamefont{Odijk}(1993{\natexlab{a}})}]{odijk1993a}
\bibinfo{author}{\bibfnamefont{T.}~\bibnamefont{Odijk}},
  \bibinfo{journal}{Biophys. Chem.} \textbf{\bibinfo{volume}{\textbf{46}}},
  \bibinfo{pages}{69} (\bibinfo{year}{1993}{\natexlab{a}}).

\bibitem[{\citenamefont{Odijk}(1993{\natexlab{b}})}]{odijk1993b}
\bibinfo{author}{\bibfnamefont{T.}~\bibnamefont{Odijk}},
  \bibinfo{journal}{Europhys. Lett.} \textbf{\bibinfo{volume}{\textbf{24}}},
  \bibinfo{pages}{177} (\bibinfo{year}{1993}{\natexlab{b}}).

\bibitem[{\citenamefont{Pereira and Williams}(2000)}]{pereira2000}
\bibinfo{author}{\bibfnamefont{G.~G.} \bibnamefont{Pereira}} \bibnamefont{and}
  \bibinfo{author}{\bibfnamefont{D.~R.~M.} \bibnamefont{Williams}},
  \bibinfo{journal}{Europhys. Lett.} \textbf{\bibinfo{volume}{\textbf{50}}},
  \bibinfo{pages}{559} (\bibinfo{year}{2000}).

\bibitem[{\citenamefont{Wulff}(1901)}]{wulff1901}
\bibinfo{author}{\bibfnamefont{G.}~\bibnamefont{Wulff}},
  \bibinfo{journal}{Zeitschrift f{\"u}r Kristallographie}
  \textbf{\bibinfo{volume}{\textbf{34}}}, \bibinfo{pages}{449}
  (\bibinfo{year}{1901}).

\bibitem[{\citenamefont{Love}(1944)}]{love1944treatise}
\bibinfo{author}{\bibfnamefont{A.~E.~H.} \bibnamefont{Love}},
  \emph{\bibinfo{title}{A Treatise on the Mathematical Theory of Elasticity}}
  (\bibinfo{publisher}{Dover Publications}, \bibinfo{address}{New York},
  \bibinfo{year}{1944}).

\bibitem[{\citenamefont{Gittes}(1994)}]{gittesthesis}
\bibinfo{author}{\bibfnamefont{F.}~\bibnamefont{Gittes}}, Ph.D. thesis,
  \bibinfo{school}{University of Washington} (\bibinfo{year}{1994}).

\bibitem[{\citenamefont{Schiessel et~al.}(2000)\citenamefont{Schiessel,
  Rudnick, Bruinsma, and Gelbart}}]{schiessel2000}
\bibinfo{author}{\bibfnamefont{H.}~\bibnamefont{Schiessel}},
  \bibinfo{author}{\bibfnamefont{J.}~\bibnamefont{Rudnick}},
  \bibinfo{author}{\bibfnamefont{R.}~\bibnamefont{Bruinsma}}, \bibnamefont{and}
  \bibinfo{author}{\bibfnamefont{W.~M.} \bibnamefont{Gelbart}},
  \bibinfo{journal}{Europhys. Lett.} \textbf{\bibinfo{volume}{\textbf{51}}},
  \bibinfo{pages}{237} (\bibinfo{year}{2000}).

\end{thebibliography}
\end{document}